\documentclass[10pt,preprint]{aastex}

\newcommand{\bu}{ \mathbf{u} }
\newcommand{\divu}{ \mathbf{\nabla} \cdot \bu }

\shorttitle{CRASH: AN AMR CODE FOR RADHYDRO}
\shortauthors{van der Holst et al.}

\begin{document}

\title{CRASH: A BLOCK-ADAPTIVE-MESH CODE FOR RADIATIVE SHOCK HYDRODYNAMICS
-- IMPLEMENTATION AND VERIFICATION}

\author{B. van der Holst\altaffilmark{1},
  G.   T\'oth\altaffilmark{1},
  I.V. Sokolov\altaffilmark{1},
  K.G. Powell\altaffilmark{1},
  J.P. Holloway\altaffilmark{1},
  E.S. Myra\altaffilmark{1},
  Q.   Stout\altaffilmark{1},
  M.L. Adams\altaffilmark{2},
  J.E. Morel\altaffilmark{2},
  R.P. Drake\altaffilmark{1}}

\altaffiltext{1}{University of Michigan, Ann Arbor, MI 48109, USA}
\altaffiltext{2}{Texas A\&M University, College Station, TX, USA}

\begin{abstract}
We describe the CRASH (Center for Radiative Shock Hydrodynamics) code, a
block adaptive mesh code for multi-material radiation hydrodynamics. The
implementation solves the radiation diffusion model with the gray or
multigroup method and uses a flux limited diffusion approximation to recover
the free-streaming limit. The electrons and ions are allowed to have different
temperatures and we include a flux limited electron heat conduction. The
radiation hydrodynamic equations are solved in the Eulerian frame by means of a
conservative finite volume discretization in either one, two, or
three-dimensional slab geometry or in two-dimensional cylindrical symmetry.
An operator split method is used to solve these equations in three substeps:
(1) solve the hydrodynamic equations with shock-capturing schemes,
(2) a linear advection of the radiation in frequency-logarithm space, and
(3) an implicit solve of the stiff radiation diffusion, heat conduction, and
energy exchange. We present a suite of verification test problems to
demonstrate the accuracy and performance of the algorithms. The CRASH code is
an extension of the Block-Adaptive Tree Solarwind Roe Upwind Scheme
(BATS-R-US) code with this new radiation transfer and heat conduction library
and equation-of-state and multigroup opacity solvers. Both CRASH and BATS-R-US
are part of the publicly available Space Weather Modeling Framework (SWMF).

\end{abstract}


\section{INTRODUCTION}

As photons travel through matter, the radiation field experiences changes due
to net total emission, absorption, and scattering,
see for instance \citet{mihalas1984,pomraming2005,drake2006}. At high
enough energy density the radiation will heat and accelerate the plasma.
This coupled system is radiation hydrodynamics. The radiation
can at a fundamental level be described by the time evolution of the spectral
radiation intensity $I_\nu({\bf r},t,{\bf n},\nu)$, which is the
radiation energy per unit area, per unit solid angle in the direction of
photon propagation ${\bf n}$, per unit interval of photon frequency $\nu$,
and per unit time interval. Several method have been developed to solve the
radiation field in various degree of physics fidelity.

In Monte Carlo radiative transfer methods, the radiation is statistically
evaluated. Small photon packets are created with their energy and propagation
direction statistically selected. The packets are propagated through matter
using the radiation transfer equation \citep{nayakshin2009,maselli2009,
baek2009}. Characteristic methods use integration along rays of various
lengths to solve for the angular structure of the radiation transport. A
recent conservative, causal ray-tracing method was developed and combined
with a short characteristic ray-tracing for the transfer calculations of
ionizing radiation \citep{mellema2006}. Solar surface magneto-convection
simulations are increasingly realistic
and use a three-dimensional, non-gray, approximate local thermodynamic
equilibrium (LTE), radiative transfer for the heating and cooling of plasma.
These simulations are typically formulated for four frequency bins in the
radiative transport equation \citep{vogler2005,stein2007,martinez2009}.

For some applications, simplifications to the radiation transfer can be made
by calculating moments of the radiation intensity over the solid angle
$\Omega$. The spectral radiation energy and the spectral radiation energy flux
are defined by the 0th and 1st moments as
\begin{equation}
  E_\nu({\bf r},t,\nu) = \frac{1}{c}\int_{4\pi} I_\nu({\bf r},t,{\bf n},\nu)
  d\Omega, \qquad
  {\bf F}_\nu({\bf r},t,\nu) = \int_{4\pi} {\bf n} I_\nu({\bf r},t,{\bf n},\nu)
  d\Omega.
\end{equation}
In addition, the spectral radiation pressure tensor ${\bf P}_\nu$ is defined
by the second moment
\begin{equation}
  {\bf P}_\nu({\bf r},t,\nu) = \frac{1}{c} \int_{4\pi}
         {\bf n}{\bf n} I_\nu({\bf r},t,{\bf n},\nu)d\Omega.
         \label{eq:pressuretensor}
\end{equation}
A whole class of radiation transfer models are based on solving the
corresponding radiation moment equations, using a closure relation between
the spectral pressure tensor and the spectral intensity
\citep{mihalas1984,pomraming2005,drake2006}.

A radiation-hydrodynamics code based on variable Eddington tensor (VET)
methods \citep{stone1992} can still capture the angular structure of the
radiation field by relating the spectral radiation pressure tensor to
the spectral radiation intensity and the method is applicable for both the
optically thin and thick regime. Optically thin versions of the VET method
have been used in the context of cosmological reionization \citep{petkova2009}.

Further simplification assumes that the radiation pressure is isotropic
and proportional to the radiation energy. This is the diffusion approximation.
Several codes have been developed using this approximation. HYDRA
\citep{marinak2001} is an arbitrary Lagrange Eulerian code for 2D and 3D
radiation hydrodynamics. The radiation transfer model is based upon either flux
limited multigroup or implicit Monte Carlo radiation transport. The Eulerian
code RAGE \citep{gittings2008} uses a cell-based adaptive mesh refinement to
achieve resolved radiative hydrodynamic flows. HYADES \citep{larsen1994} solve
the radiation hydrodynamic equations on a Lagrangian mesh, while CALE
\citep{barton1985} can use either a fixed Eulerian mesh, an embedded
Lagrangian mesh, or a partially embedded, partially
remapped mesh. Our newly developed radiation-hydrodynamic solver uses an
Eulerian grid together with a block-based adaptive mesh refinement strategy.

We limit the discussion of the radiation hydrodynamics implementation in CRASH
to plasmas in the absence of magnetic field. Most of the description in this
paper can,
however, easily be extended to magnetohydrodynamic (MHD) plasmas as well.
Indeed, since the CRASH code is essentially the magnetohydrodynamic BATS-R-US
code \citep{powell1999,toth2010} extended with libraries containing radiation
transport, equation-of-state (EOS), and opacity solvers, the implementation
for the coupling between the radiation field and MHD plasmas is readily
available. The CRASH code uses the recently developed block adaptive
tree library (BATL, \citet{toth2010}). Here we will focus on the radiation
implementation. Both the CRASH and BATS-R-US codes are publicly available as
part of the Space Weather Modeling Framework (SWMF, \citet{toth2005}) or
can be used as stand-alone codes.

In the following, Section \ref{sec:radhydro} introduces the radiation
hydrodynamic equations for multi-material plasmas, in a form general enough to
apply at high energy density.
Section \ref{sec:method} describes the numerical algorithms to solve these
equations. Next, in Section \ref{sec:verification} verification tests, are
presented for radiation and electron heat conduction on non-uniform meshes in
1D, 2D, and 3D slab geometry and in axially symmetric ($rz$) geometry. We also
show a full system multi-material radiation hydrodynamic simulation on an
adaptively refined mesh and demonstrate good scaling up to 1000 processors.
The paper is summarized in Section \ref{sec:summary}.

\section{EQUATIONS OF RADIATION HYDRODYNAMICS IN DENSE PLASMAS}
\label{sec:radhydro}

The equations of radiation hydrodynamics describe the time evolution of
both matter and radiation. For the applications that supported the work
reported here, the code must be able to model matter as a high energy density
plasma that is in LTE so that the population of all atomic and ion states can
be obtained from statistical physics (see for instance \citet{landau1980}).
We allow for multiple materials throughout the spatial domain of interest,
but restrict the analysis to plasma flows that are far from relativistic. The
materials can be heated to sufficiently high temperatures so that they can
ionize and create free electrons, introducing the need for a time evolution
equation for the electron energy density. The electrons transfer heat by
thermal heat conduction and emit and absorb photon radiation. The radiation
model discussed in this paper is non-equilibrium diffusion, in which the
electron and radiation temperature can be different. We approximate the
radiation transfer with a gray or multigroup flux limited diffusion (FLD).
This model is also of interest for application to a number of astrophysical
problems.

In the following subsections, we will describe the radiative transfer equations
for the evolution of the multigroup radiation energy densities (Section
\ref{sec:radtrans}) in the FLD approximation (Section \ref{sec:fld}).
The coupling of the radiation field to the two species hydrodynamic equations
of electrons and ions are discussed in Section \ref{sec:hydroeq}. In Section
\ref{sec:levelset}, the method for tracking the different materials is
treated, while the lookup tables used for of the EOS and opacities are
mentioned in Section \ref{sec:eos}.

\subsection{Radiation Transport}
\label{sec:radtrans}

In this section, we will build up the form of the radiation transport
in the multigroup diffusion approximation that is used for the implementation
in the CRASH code. The spectral pressure tensor, equation
(\ref{eq:pressuretensor}), is often approximated in the form
\citep{mihalas1984}
\begin{equation}
  {\bf P}_\nu({\bf r},t,\nu) = E_\nu {\bf T}_\nu,
\end{equation}
where
\begin{equation}
  {\bf T}_\nu({\bf r},t,\nu) = \frac{1}{2}(1-\chi_\nu){\bf I}
  + \frac{1}{2}(3\chi_\nu-1)\frac{{\bf F}_\nu{\bf F}_\nu}{|{\bf F}_\nu|^2},
\end{equation}
is the spectral Eddington tensor, $\chi_\nu$ is the Eddington factor, and
${\bf I}$ is the identity matrix. The second term on the right hand side is a
dyad constructed from the direction of the spectral radiation flux.
The pressure tensor can be used to arrive at a time evolution equation for the
solid angle integrated spectral radiation energy \citep{buchler1983}
\begin{equation}
  \frac{\partial E_\nu}{\partial t}+\nabla\cdot(E_\nu \bu)
  -\nu\frac{\partial}{\partial \nu}
  ({\bf P}_\nu:\nabla{\bf u}) =
  {\rm diffusion + emission - absorption}, \label{eq:buchlerform}
\end{equation}
which contains the plasma velocity $\bu$ of the background plasma.
Here the colon denotes the contraction of the two tensors ${\bf P}_\nu$ and
$\nabla\bu$. The processes described by the symbolic terms on the right hand
side of equation (\ref{eq:buchlerform}) will be described below.

Setting the Eddington factor $\chi_\nu=1/3$ corresponds to the radiation
diffusion model. In this case the radiation is assumed to be effectively
isotropic and the spectral radiation pressure can be described by the scalar
\begin{equation}
  p_\nu = \frac{1}{3}E_\nu=(\gamma_r - 1)E_\nu, \label{eq:pressure}
\end{equation}
where we have introduced the adiabatic index of the radiation field, which
in this case has the relativistic value $\gamma_r=4/3$. The time evolution for
the spectral energy density can then be simplified to
\begin{equation}\label{eq:mg1}
  \frac{\partial E_\nu}{\partial t}+\nabla\cdot(E_\nu \bu)
  -(\gamma_r-1)(\divu)\nu
  \frac{\partial E_\nu}{\partial \nu} =
       {\rm diffusion + emission - absorption}.
\end{equation}
The second and third terms on the left hand side of equation (\ref{eq:mg1})
express the change in the spectral energy density due to the advection and
compression of the background plasma, which moves with the velocity
$\bu$, as well as the frequency shift due to compression. In the free-streaming
limit where the radiation hardly interacts with the matter, $\chi_\nu$ will
approach one. In this paper we will keep $\chi_\nu=1/3$ and at the same time
use a flux limited diffusion for the free-streaming regime whenever needed
(see Section \ref{sec:fld}).

The set of equations for the spectral energy density (\ref{eq:mg1}) still
consists of an infinite amount of equations, one for each frequency.
A finite set of governing equations to describe the radiation transport in the
multigroup diffusion approximation is obtained when we choose a set of
frequency groups. Here we enumerate groups with the index, $g=1,\ldots, G$.
The interval of the photon frequencies, relating to the $g$th group is denoted
as $[\nu_{g-1/2},\nu_{g+1/2}]$. A discrete set of group energy densities,
$E_g$, is introduced in terms of the integrals of the spectral energy density
of the frequency group interval:
\begin{equation}\label{eq:mg2}
  E_g = \int_{\nu_{g-1/2}}^{\nu_{g+1/2}}
  {E_\nu d\nu}. 
\end{equation}
Now we can integrate equation (\ref{eq:mg1}) to arrive at the desired set of
the multigroup equations:
\begin{eqnarray}
  \frac{\partial E_g}{\partial t} &+& \nabla\cdot(E_g \bu)+
  (\gamma_r-1)E_g\divu - (\gamma_r-1)(\divu)
    \int_{\nu_{g-1/2}}^{\nu_{g+1/2}}
    \frac{\partial(\nu E_\nu)}{\partial\nu}d\nu \nonumber\\
  &=& \int_{\nu_{g-1/2}}^{\nu_{g+1/2}}
{\rm (diffusion + emission - absorption)}d\nu.\label{eq:multigroup}
\end{eqnarray}
The fourth term on the left hand side is a frequency shift due to the plasma
compression. This term is essentially a conservative advection along the
frequency axis.

In the context of the multi-group radiation diffusion, a discussion about
the stimulated emission is not less important than LTE.
Excellent accounts on the stimulated emission exist in the literature, see
for instance \cite{zeldovich2002}. Here, we merely sumarize how the stimulated
emission modifies the absorption opacity $\kappa_\nu^a$ obtained from,
e.g., opacity tables. This is important when dealing with externally supplied
opacity tables, since the CRASH code assumes that the absorption opacities
are corrected. Integrating the total absorption and emission over
all directions and summing up the two polarizations of the photons, the
following expression can be derived for the emission and absorption
\begin{equation}
  {\rm emission} - {\rm absorption} = c{\kappa_\nu^a}'
  \left( B_\nu - E_\nu \right),
\end{equation}
where the effective absorption coefficient, ${\kappa_\nu^a}'$, is introduced
to account for the correction due to stimulated emission:
\begin{equation}
  {\kappa_\nu^a}' = \kappa_\nu^a \left( 1 - \exp \left[
    -\frac{\varepsilon}{k_B T_e} \right] \right),
\end{equation}
in which $\varepsilon=h\nu$ is the photon energy, $k_B$ is the Boltzmann
constant, and $T_e$ is the electron temperature. We also introduced the
spectral energy density distribution of the black body radiation
(the Planckian)
\begin{equation}
  B_\nu = \frac{8\pi}{h^3 c^3} \frac{\varepsilon^3}{\exp[ \varepsilon/
      (k_B T_e)] - 1}.\label{eq:planck}
\end{equation}
The total energy density in the Planck spectrum equals
$B=\int_0^\infty d\nu B_\nu = a T_e^4$, where $a=8\pi^5k_B^4/(15h^3c^3)$
is the radiation constant.

We use the standard definition of the group Planck mean opacity $\kappa_{Pg}$
and group Rosseland mean opacity $\kappa_{Rg}$ \citep{mihalas1984}
\begin{equation}
  \kappa_{Pg} = \frac{\int_{\nu_{g-1/2}}^{\nu_{g+1/2}}d\nu{\kappa_\nu^a}'B_\nu}
  {B_g}, \qquad
  \kappa_{Rg} = \frac{\frac{\partial B_g}{\partial T_e}}
  {\int_{\nu_{g-1/2}}^{\nu_{g+1/2}}d\nu \frac{1}{\kappa_\nu^t}
    \frac{\partial B_\nu}{\partial T_e}}, \qquad
  B_g= \int_{\nu_{g-1/2}}^{\nu_{g+1/2}} d\nu B_\nu
\end{equation}
in which $\kappa_\nu^t$ is the spectral total opacity. The  right hand side
of equation (\ref{eq:multigroup}) can now be written as
(see for instance \cite{mihalas1984,pomraming2005})
\begin{eqnarray}
  \frac{\partial E_g}{\partial t} &+& \nabla\cdot(E_g\bu)
  + (\gamma_r - 1)E_g\divu
  - (\gamma_r - 1)\divu \int_{\nu_{g-1/2}}^{\nu_{g+1/2}}
  \frac{\partial(\nu E_\nu)}{\partial\nu}d\nu \nonumber \\
  &=& \nabla\cdot\left( D_g\nabla E_g\right) + \sigma_g (B_g-E_g),
  \label{eq:groupenergy}
\end{eqnarray}
where $D_g=c/(3\kappa_{Rg})$ is the radiation diffusion coefficient for
radiation group $g$. The absorption and emission uses the coefficient
$\sigma_g=c\kappa_{Pg}$. These group mean opacities are either supplied by
lookup tables or by an opacity solver.

In a single group approximation (gray diffusion), the spectral energy density
is integrated over all photon frequencies and the total radiation energy
density is obtained by
\begin{equation}\label{eq:mg0}
  E_r({\bf r},t) = \int_0^\infty{E_\nu d\nu}.
\end{equation}
This amounts to summing up all groups $E_r = \sum_g E_g$. The gray radiation
diffusion equation can be derived as (see for instance
\cite{mihalas1984,pomraming2005,drake2006})
\begin{equation}
  \frac{\partial E_r}{\partial t} + \nabla\cdot(E_r\bu)
  + (\gamma_r - 1)E_r\divu
  = \nabla\cdot\left( D_r\nabla E_r\right) + \sigma_r (B-E_r),
  \label{eq:grayenergy}
\end{equation}
where the diffusion coefficient $D_r$ is now defined by the single group
Rosseland mean opacity $\kappa_R$ as $D_r = c/(3\kappa_{R})$, and the
absorption coefficent $\sigma_r$ is defined in terms of the single group
Planck mean opacity $\kappa_P$ as $\sigma_r = c\kappa_P$.

\subsection{Hydrodynamics}\label{sec:hydroeq}

In the CRASH code, a single fluid description is used, so that all of the
atomic and ionic species as well as the electrons move with the same bulk
velocity $\bu$. The conservation of mass
\begin{equation}
  \frac{\partial\rho}{\partial t} + \nabla\cdot(\rho\bu) = 0,
  \label{eq:density}
\end{equation}
provides the time evolution of the mass density $\rho$ of all the materials
in the simulation. The plasma velocity is obtained from the conservation of
momentum
\begin{equation}
  \frac{\partial\rho\bu}{\partial t} + \nabla\cdot\left[
    \rho\bu\bu + I(p+p_r)\right] = 0. \label{eq:momentum}
\end{equation}
The total plasma pressure is the sum of the ion and electron pressures:
$p = p_i + p_e$. The net force of the radiation on the plasma is given by the
gradient of the total radiation pressure $-\nabla p_r$, where the total
radiation pressure is obtained from the group radiation energies:
$p_r = (\gamma_r -1)\sum E_g$.

In a high density plasma, the electrons are very strongly coupled to the ions
by collisions. However, for higher temperatures, the electrons and ions get
increasingly decoupled. At a shock front, where ions are preferentially
heated by the shock
wave, the electrons and ions are no longer in temperature equilibrium. Ion
energy is transfered to the electrons by collisions, while the electrons will
in turn radiate energy. We therefore solve separate equations for the
ion/atomic internal energy density $E_i$ and the electron internal energy
density $E_e$:
\begin{eqnarray}
  \frac{\partial E_i}{\partial t} &+& \nabla\cdot(E_i\bu)+p_i\divu =
  \sigma_{ie} (T_e - T_i), \label{eq:ionenergy} \\
  \frac{\partial E_e}{\partial t} &+& \nabla\cdot(E_e\bu)+p_e\divu =
  \nabla\cdot(C_e\nabla T_e) + \sigma_{ie} (T_i - T_e)
  + \sum_{g=1}^G \sigma_g (E_g-B_g). \label{eq:electronenergy}
\end{eqnarray}
The coupling coefficient $\sigma_{ie}=n_a k_B/\tau_{ie}$ in the collisional
energy exchange between the electrons and ions depends on the relaxation time
$\tau_{ie}(T_e, n_a, m)$ and the atomic number density $n_a$.
The energy transfer depends also on the difference between the
ion temperature $T_i$ and the electron temperature $T_e$.
In equation (\ref{eq:electronenergy}), we have included the electron thermal
heat conduction with conductivity $C_e(T_e, n_a, m)$. Since the electrons are
the species that are responsible for the radiation absorption and
emission, the energy exchange between the electrons and the radiation groups
is added to equation (\ref{eq:electronenergy}).

For the development of the numerical schemes in Section \ref{sec:method} we
will use an equation for the conservation of the total energy density
\begin{equation}
  e =  \frac{\rho u^2}{2} + E_i + E_e + \sum_{g=1}^G E_g,
  \label{eq:totalenergy}  
\end{equation}
instead of the equation for the ion internal energy (\ref{eq:ionenergy}).
This is especially important in regions of the computational domain where
hydrodynamic shocks can occur, so that we can recover the correct jump
conditions. The conservation of the total energy can be derived from equations
(\ref{eq:groupenergy}) and (\ref{eq:density})--(\ref{eq:electronenergy}) as
\begin{equation}
  \frac{\partial e}{\partial t} + \nabla\cdot\left[ (e+p+p_r)\bu\right]
  = \nabla\cdot(C_e\nabla T_e) +
  \sum_{g=1}^G \nabla\cdot\left( D_g\nabla E_g\right).
  \label{eq:energycons}
\end{equation}
The frequency shift term in equation (\ref{eq:groupenergy}) due to
the plasma compression does not show up in the conservation of the
total energy if we use energy conserving boundary conditions at the end
points of the frequency domain, i.e. at $\nu=0$ and $\nu=\infty$ in the
analytical description or at the end points of the numerically truncated
finite domain.

\subsection{Level Sets and Material Identification}\label{sec:levelset}

In many of the CRASH applications, we need a procedure to distinguish between
different materials. We assume that the materials do not mix, but differ from
each other by their properties such as the equation of state and opacities.
If we use $M$ different materials, then we can define for each material
$m=1,\ldots, M$ the level set function $d_m({\bf r}, t)$ that is
initially set to zero at the material interface, while positive inside the
material region and negative outside. Generally, we use a smooth and signed 
distance function in the intial state. At later times, the location of
material $m$ is obtained by means of a simple advection equation
\begin{equation}
  \frac{\partial d_m}{\partial t} + \nabla\cdot( d_m \bu ) = d_m\divu.
  \label{eq:levelset}
\end{equation}
For any given point in space and time, we can determine what the material is,
since analytically only one of the level set functions $d_m$ can be positive at
any given point. Numerical errors will create regions where this is not true.
In practice we take the largest $d_m$. This is a simple approximation, we may
explore more sophisticated approaches in the future. The
number of material levels $M$ can be configured at compile time.

\subsection{Equation of State and Opacities}\label{sec:eos}

We have implemented EOS solvers and a code
to calculate the frequency averaged group opacities. The implemention will be
reported elsewhere, but mention that in the EOS and opacity solver
the temperature is assumed to be well below the relativistic values:
$T \ll 10^5$\,eV. The non-relativistic speed of motion is also assumed while
simplifying the radiation transport equation and to neglect the relativistic
corrections for opacities. In this paper, we will assume that all necessary
quantities are calculated and stored in lookup tables. Our EOS solver assumes
that the corrections associated with ionization, excitation, and Coulomb
interactions of the partially ionized ion-electron plasma are all added to
the energy of the electron gas and to the electron pressure. This is possible
since those corrections are controlled by the electron temperature. The
ion internal energy density, ion pressure, and ion specific heat in an
isochoric process per unit of volume are simply
\begin{equation}
  p_i = n_ak_BT_i, \qquad E_i = \frac{p_i}{\gamma -1},
  \qquad C_{Vi} = \left( \frac{\partial E_i}{\partial T_i} \right)_\rho
  = \frac{n_ak_B}{\gamma -1},
\end{equation}
which are due to the contributions due to ion translational motions for which
$\gamma=5/3$.

The relations among the electron internal energy density, pressure, density,
and temperature are known as the EOS. To solve these relations
is usually complex and time consuming. We therefore store these relations in
invertible lookup tables. For each material $m$, our EOS tables have the
logarithmic lookup arguments $(\log T_e, \log n_a)$.
The list of quantities stored in these tables is indicated in Table
\ref{table:eos}. These lookup tables are populated with quantities that are
needed for both single temperature and two temperature simulations.
For two-temperature plasma simulations, we will need $p_e$, $E_e$, the electron
specific heat $C_{Ve}$, the electron speed of sound gamma $\gamma_{S_e}$.
For convenience we added the total pressure $p=p_e+p_i$, total internal energy
density $E = E_i + E_e$, single temperature specific heat $C_V$, and the
single temperature speed of sound gamma $\gamma_S$, which can be used in single
temperature simulations. We use high enough table resolutions so that it
is sufficient to use a bilinear interpolation in the lookup arguments.
If $p_e$ or $E_e$ (or $p$ and $E$ in single temperature mode) are known on
entry of the lookup instead of $T_e$, we do a binary search in the table to
find the appropriate electron temperature. The latter only works as long as the
necessary thermodynamic derivatives are sign definite, i.e. the table is
invertible. Other thermodynamic quantities that are needed, but not stored in
these lookup tables, can be derived. For example, the electron density can be
obtained from the mean ionization $n_e = n_a \overline{Z}$.

In addition, we have lookup tables for the averaged multigroup opacities.
These tables are either constructed internally for a given frequency range,
number of groups, and the selected materials, or externally supplied. For any
material $m$, the logarithmic lookup arguments are $(\log \rho, \log T_e)$.
The stored quantities, see Table \ref{table:opacity}, are the specific
Rosseland mean opacity $\kappa_{Rg}/\rho$ and the specific Planck mean opacity
$\kappa_{Pg}/\rho$ for all groups $g=1,\ldots, G$ that are used during a
simulation. The Planck opacities are assumed to be corrected for the
stimulated emission, as discussed in Section \ref{sec:radtrans}. The groups are
always assumed to be logarithmically distributed in the frequency space.

\subsection{Flux Limited Diffusion}\label{sec:fld}

Radiation diffusion theory can transport energy too fast in the optically
thin free streaming limit. In the diffusion limit, the radiation diffusion
flux for each group follows Fick's law ${\bf F}_g = -D_g\nabla E_g$, where
the diffusion coefficient $D_g$ depends on the Rosseland mean opacity
$\kappa_{Rg}$ for the group $g$ via $D_g = c/(3\kappa_{Rg})$. This flux is
however not bounded. In the optically thin free-streaming limit, the magnitude
of the radiation flux can be at most $cE_g$ in order to maintain causality.
Various flux limiters exist in the literature, see for instance
\citet{minerbo1978,lund1980,levermore1981}, that ensure that the diffusion
flux is limited by this free streaming flux. We implemented the so-called
square-root flux limiter to obtain the correct progation speed in the
optically thin regime \citep{morel2000}. For this flux limiter, the diffusion
coefficient is rewritten as
\begin{equation}
  D_g = \frac{c}{\sqrt{(3\kappa_{Rg})^2 + \frac{|\nabla E_g|^2}{E_g^2}}}.
\end{equation}
In the limit that the radiation length scale $L_R = E_g/|\nabla E_g|$ is large,
the diffusive limit is recovered. For a small radiation length scale,
$D_g = c|E_g|/|\nabla E_g|$ and the radiation propagates with the speed of
light.

Similarly, we implemented the option to limit the electron thermal
heat flux (see \citet{drake2006} for more details on electron flux
limiters). The classical Spitzer-Harm formula for the collisional electron
conductivity is proportional to $T_e^{5/2}/\overline{Z^2}$, where
$\overline{Z^2}$ is the mean square ionization of the used material.
The collisional model is only valid
when the temperature scale length $L_T = T_e/|\nabla T_e|$ is much larger than
the collisional mean free path of the electrons $\lambda_{mfp}$. When the
temperature scale length is only a few $\lambda_{mfp}$ or smaller, this
description breaks down. This may for instance happen in
laser-irradiated plasmas. One could in that case find the heat flux from
solving the Fokker-Planck equation for the electrons, but this is
computationally expensive. Instead, we use a simplified model to limit the
electron heat flux. A free-streaming heat flux can
be defined as the thermal energy density in the plasma transported at some
characteristic thermal velocity: $F_{FS} = n_ek_BT_ev_{th}$, where
$v_{th}=\sqrt{k_BT_e/m_e}$. For practical applications, the maximum heat
transport is usually only a fraction of this free-streaming flux:
${\bf F} = -(fF_{FS}/|\nabla T_e|) \nabla Te$, where $f$ is the so-called
flux limiter. This heat flux model is the threshold model and is also used
in other radhydro packages, such as HYADES \citep{larsen1994}.
The flux-limited heat flux can now be defined as
\begin{equation}
  {\bf F} = -\min\left(C_e,\frac{fF_{FS}}{|\nabla T_e|}\right)\nabla T_e.
\end{equation}
The flux limiter $f$ is an input parameter and can be tuned to let the
simulated results better fit reality.

\section{THE NUMERICAL METHOD}\label{sec:method}

In this section, we describe the discretization of the set of multi-material,
radiation hydrodynamics equations for the density (\ref{eq:density}), momentum
(\ref{eq:momentum}), total energy (\ref{eq:energycons}), electron internal
energies (\ref{eq:electronenergy}), radiation group energy
(\ref{eq:groupenergy}), and material level set functions (\ref{eq:levelset}).
The equations are time integrated using an operator split method to solve the
equations in substeps. Formally, we may write this system as
\begin{equation}
  \frac{\partial {\bf U}}{\partial t} = {\bf R}_{\mbox{hydro}}({\bf U})
  + {\bf R}_{\mbox{frequency}}({\bf U}) + {\bf R}_{\mbox{diffusion}}({\bf U}),
\end{equation}
where ${\bf U}$ is the vector of state variables. We have split the right
hand sides of the equations into three parts and time advance the equations
with an operator splitting method in the following order:
(1) The right hand side ${\bf R}_{\mbox{hydro}}$ describes
the advection and pressure contributions (Section \ref{sec:hydro}).
This part is essentially the ideal hydrodynamic equations augmented with the
advection and compression of the radiation energy, the electron internal
energy, and the material level sets.
(2) The right hand side ${\bf R}_{\mbox{frequency}}$ is the advection of the
radiation field in frequency space (Section \ref{sec:frequency}).
(3) The right hand side ${\bf R}_{\mbox{diffusion}}$ takes care of the
diffusion and energy exchange terms, which we will solve with an implicit
scheme (Section \ref{sec:implicit}). The operator splitting is not unique.
Instead of splitting the hydrodynamic advection operator and the extra
advance operator for the frequency advection, one may attempt to discretize
the frequency advection flux as an extra flux for the control volume of the
four-dimensional $(x,y,z,\nu)$ space. However, since the CRASH code is built
around the existing BATS-R-US code in 1D, 2D, and 3D, we opted for splitting
the frequency advection from the hydro update. The boundary conditions are
treated in Section \ref{sec:boundary}.

\subsection{Hydro Solve}\label{sec:hydro}

In the first step of the operator splitting, we update the hydrodynamic
equations including the advection and compression of the radiation energy
density, electron internal energy density and the level sets. We have
implemented two variants to solve the hydrodynamic equations: (1) using
conservation of the total energy (Section \ref{sec:conservative}) and
(2) a non-conservative pressure formulation (Section
\ref{sec:nonconservative}). We can combine the two discretizations in a
hybrid manner a simulation.

\subsubsection{Conservative}\label{sec:conservative}

We have implemented several hydrodynamic shock-capturing schemes in the CRASH
code: the HLLE scheme \citep{harten1983,einfeldt1991}, the Rusanov
scheme \citep{yee1989}, and a Godunov scheme \citep{godunov1959} with an exact
Riemann solver. In this section, we will explain how we generalized the HLLE
scheme for our system of equations that includes radiation, level sets, and an
EOS. The other hydrodynamic schemes can be generalized in a similar fashion.

Typical hydrodynamic solvers in the literature assume constant $\gamma$. Our
problem is to generalize the constant $\gamma$ hydro solvers for
the case of spatially varying polytropic index, $\gamma_e$, due to ionization,
excitation and Coulomb interactions. A method that is applicable to all the
aforementioned, constant $\gamma$, hydrodynamic shock-capturing schemes is to
split the electron internal energy $E_e$ density as the sum of an ideal
(translational) energy part $p_e/(\gamma -1)$ and an extra internal energy
density $E_X$. Similarly, we can define an ideal total energy density
\begin{equation}
  e_I =  \frac{\rho u^2}{2} + \frac{p_i+p_e}{\gamma-1} + \sum_{g=1}^G E_g,
\end{equation}
which is related to the total energy density by $e = e_I + E_X$.
We will time advance $p_e$ with the ideal electron pressure equation and
$E_X$ by a conservative advection equation, and then apply a correction step
as described below.

The time update with the operator ${\bf R}_{\mbox{hydro}}$ solves the following
equations:
\begin{eqnarray}
  \frac{\partial\rho}{\partial t} &+& \nabla\cdot(\rho\bu) = 0,
  \label{eq:hddens}\\
  \frac{\partial\rho\bu}{\partial t} &+& \nabla\cdot\left[
    \rho\bu\bu + I(p+p_r)\right] = 0, \\
  \frac{\partial e_I}{\partial t} &+& \nabla\cdot\left[ (e_I+p+p_r)\bu\right]
  = 0,\label{eq:hdenergy}\\
  \frac{1}{\gamma-1}\frac{\partial p_e}{\partial t} &+&
  \frac{1}{\gamma-1}\nabla\cdot(p_e\bu)+p_e\divu
  = 0, \label{eq:hdelectronenergy} \\
  \frac{\partial E_X}{\partial t} &+& \nabla\cdot \left[ E_X \bu \right]
  =0, \label{eq:hdextra} \\
  \frac{\partial E_g}{\partial t} &+& \nabla\cdot(E_g\bu)
  + (\gamma_r - 1)E_g\divu  = 0, \label{eq:hdradiation} \\
  \frac{\partial d_m}{\partial t} &+& \nabla\cdot( d_m \bu ) - d_m\divu =0,
  \label{eq:hdlevel}
\end{eqnarray}
where the frequency advection, diffusion, and energy exchange terms are
ommitted in this first operator step. After each time advance from time
$t^n$ to time $t^{n+1}$, we have to correct $e$, $e_I$, $p_e$, and $E_X$. We
denote the uncorrected variables with a superscript $*$, then we recover at
time level $n+1$ the true electron internal energy $E_e^{n+1}$ and the
true total energy density $e^{n+1}$ by
\begin{eqnarray}
  E_e^{n+1} &=& \frac{p_e^*}{\gamma-1} + E_X^*, \label{eq:fixstart}\\
  e^{n+1} &=& e_I^* + E_X^*.
\end{eqnarray}
Since both $e_I$ and $E_X$ follow a conservation law, the total energy
density $e$ is also conserved.
The true electron pressure is recovered from the updated electron internal
energy and mass density by means of the EOS:
\begin{equation}
  p_e^{n+1} = p_{\rm EOS}(\rho^{n+1}, E_e^{n+1},m),
  \end{equation}
where the function $p_{\rm EOS}$ can be either a calculated EOS or an EOS
lookup table for material $m$, determined by the level set functions
$d_m^{n+1}$ (Section \ref{sec:levelset}). The extra internal energy $E_X$ is
reset as the difference between the true electron internal energy and the
ideal electron internal energy for $\gamma=5/3$:
\begin{equation}
  E_X^{n+1} = E_e^{n+1} - \frac{p_e^{n+1}}{\gamma -1}.
\end{equation}
This is postive because the EOS state $p_{\rm EOS}$ satisfies
$E_e-p_e/(\gamma -1)\ge0$ at all times. The ideal part of the total energy
density at time level $n+1$ can now be updated as
\begin{equation}
  e_I^{n+1} = e^{n+1} - E_X^{n+1}.\label{eq:fixstop}
\end{equation}
We have now recovered $e^{n+1}$, ${e_I}^{n+1}$, $p_e^{n+1}$, and
$E_X^{n+1}$ at time $t^{n+1}$.

We time advanced the hydrodynamic equations to the time level $*$ with a
shock-capturing scheme with a constant $\gamma=5/3$. For an ideal EOS, the
speed of sound of the equations (\ref{eq:hddens})--(\ref{eq:hdradiation}) can
be derived as
\begin{equation}
  c_s = \sqrt{\frac{\gamma (p_i+p_e) + \gamma_r p_r}{\rho}},
  \label{eq:soundspeed}
\end{equation}
which includes the modifications due to the presence of the total radiation
pressure. This speed of sound will be used in the hydro scheme below.
Since the CRASH EOS solver always satisfies $E_X\ge0$ and
$\gamma_e\le 5/3$, the speed of sound for the ideal EOS is always
an upper bound for the true speed of sound.

We use shock-capturing schemes to advance the equations
(\ref{eq:hddens})--(\ref{eq:hdlevel}). In the following, we denote the
(near) conservative variables by $U = (\rho, \rho\bu, e_I, p_e, E_X, E_g, d_m)$
and let $U$ be grid cell averages in the standard finite volume sense.
If we assume for the moment a 1D grid with spacing $\Delta x$, cell center
index $i$ and cell face between cell $i$ and $i+1$ identified by half indices
$i+1/2$, then we can write the two-stage Runge-Kutta hydro update as
\begin{eqnarray}
  U_i^{n+1/2} &=& U_i^n - \frac{\Delta t}{2\Delta x}\left(
    f_{i+1/2}^n - f_{i-1/2}^n \right), \\
  U_i^{n+1} &=& U_i^n - \frac{\Delta t}{\Delta x}\left(
    f_{i+1/2}^{n+1/2} - f_{i-1/2}^{n+1/2} \right).
\end{eqnarray}
where $f$ is the numerical flux. In particular, the HLLE flux $f$ equals the
physical flux $F(U_{i+1/2}^R)$ when $c_s^+=u_i+c_s\le0$, $F(U_{i+1/2}^L)$
when $c_s^-=u_i-c_s\ge0$, and in all other cases it uses the
weighted flux
\begin{equation}
  f_{i+1/2} = \frac{c_s^+F(U_{i+1/2}^L) - c_s^-F(U_{i+1/2}^R)
    + c_s^+c_s^-(U_{i+1/2}^R - U_{i+1/2}^L)}{c_s^+ - c_s^-}.
\end{equation}
Here, the left and right cell face states are
\begin{eqnarray}
  U_{i+1/2}^L &=& U_i + \frac{1}{2}\bar{\Delta}^LU_i, \\
  U_{i+1/2}^R &=& U_{i+1} - \frac{1}{2}\bar{\Delta}^RU_{i+1}.
\end{eqnarray}
We use the generalized Koren limiter, and define the limited slopes as
\begin{eqnarray}
  \bar{\Delta}^LU_i &=& {\rm minmod}\left[ \beta (U_{i+1}-U_i),
    \beta (U_i - U_{i-1}), \frac{2U_{i+1}-U_i-U_{i-1}}{3} \right], \\
  \bar{\Delta}^RU_i &=& {\rm minmod}\left[ \beta (U_{i+1}-U_i),
    \beta (U_i - U_{i-1}), \frac{U_{i+1}-U_i-2U_{i-1}}{3} \right],
\end{eqnarray}
for the extrapolations from the left and right. This reconstruction can be
third order in smooth regions away from extrema \citep{koren1993,toth2008}.
The parameter $\beta$ can be changed between 1 and 2, but in simulations with
adaptive mesh refinement we have best experience with $\beta = 3/2$. We
generally apply the slope limiters on the primitive variables
$(\rho, \bu, p_i, p_e, E_X/\rho, E_g, d_m)$, instead of the
conservative variables. We apply the slope limiter on $E_X/\rho$ instead of
$E_X$ since $E_X/\rho$ is smoother at shocks and across material interfaces.
A multi-dimensional update is obtained by adding the fluxes for each direction
in a dimensionally unsplit manner.

After each stage of the two step Runge-Kutta, we correct for the EOS effects
via the update procedure outlined in equations
(\ref{eq:fixstart})--(\ref{eq:fixstop}).

\subsubsection{Non-Conservative Pressure Equations}
\label{sec:nonconservative}

In regions away from shocks it is sometimes more important to preserve
pressure balance than to have a shock capturing scheme that recovers the
correct jump conditions. This is especially important at material interfaces.
We therefore have implemented the option to solve the hydro part of the
pressure equations 
\begin{eqnarray}
  \frac{\partial p_i}{\partial t} &+& \nabla\cdot(p_i\bu)
  + (\gamma -1)p_i\divu = 0, \\
  \frac{\partial p_e}{\partial t} &+& \nabla\cdot(p_e\bu)
  + (\gamma_{Se} -1)p_e\divu = 0,
  \label{eq:pressuresolve}
\end{eqnarray}
instead of the equations for the total energy (\ref{eq:hdenergy}) and
the electron internal energy (\ref{eq:hdelectronenergy}).
As long as the speed of sound gamma for the electrons
\begin{equation}
  \gamma_{S_e} = \frac{\rho}{p_e}
  \left(\frac{\partial p_e}{\partial\rho}\right)_{S_e}
\end{equation}
is smaller than $\gamma=5/3$, the numerical scheme is stable. Contrary to the
energy conserving scheme, the pressure based scheme can directly include
the EOS and we no longer need the time evolution of the extra internal
energy density (\ref{eq:hdextra}). The EOS contribution in the electron
pressure equation (\ref{eq:pressuresolve}) is implemented as a source term
$-(\gamma_{Se}-\gamma)p_e\divu$ added to the ideal electron pressure equation.

To facilitate using both the shock capturing properties and the pressure
balance at the material interfaces during CRASH simulations, we have
several criteria to automatically switch between them accordingly.
One of the criteria, for instance, uses a detection of steep pressure
gradients as a shock identification. The user can select the magnitude of the
pressure gradient above which the scheme switches to the conservative energy
equations.

\subsection{Frequency Advection}\label{sec:frequency}

The set of multigroup equations (\ref{eq:groupenergy}) contain an
integral over the group photon frequencies. Performing this integration, the
frequency advection update by the ${\bf R}_{\mbox{frequency}}$ operator can
be written as
\begin{equation}
  \frac{\partial E_g}{\partial t} -(\gamma_r-1)(\nabla\cdot{\bf u})\left[
    \nu_{g+1/2} E_\nu (\nu_{g+1/2})- \nu_{g-1/2} E_\nu(\nu_{g-1/2})\right] = 0.
  \label{eq:groupbounds}
\end{equation}
These equations, however, do still depend on the unknown photon frequency
$\nu$ and the spectral radiation energy density $E_\nu$. We will now restrict
the analysis to a frequency grid that is uniformly spaced in the frequency
logarithm, i.e.,
\begin{equation}\label{eq:groupspace}
  \ln(\nu_{g+1/2}) - \ln(\nu_{g-1/2}) = \Delta(\ln \nu) = {\rm constant}.
\end{equation}
For large enough number of frequency groups $G$, the group energy $E_g$ can
then be approximated as the product of the photon frequency, spectral
radiation energy $E_\nu$, and the logarithmic group spacing $\Delta(\ln\nu)$:
\begin{equation}
  E_g = \int_{\nu_{g-1/2}}^{\nu_{g+1/2}}E_\nu d\nu =
  \int_{\ln\nu_{g-1/2}}^{\ln\nu_{g+1/2}}E_\nu\nu d(\ln\nu)
  \approx E_\nu\nu\Delta(\ln\nu).
  \label{eq:energyapprox}
\end{equation}
Using this approximation in equation (\ref{eq:groupbounds}), we obtain our
final form of the frequency advection
\begin{equation}
  \frac{\partial E_g}{\partial t} + u_\nu \frac{E_{g+1/2}- E_{g-1/2}}
       {\Delta(\ln \nu)} = 0,\label{eq:closed}
\end{equation}
where $u_\nu = -(\gamma_r-1)\divu$ is the frequency advection speed.
The values $E_{g\pm1/2}$ are to be interpolated from the mesh-centered values
$E_g$ towards the group boundaries.

The frequency advection is a conservative linear advection in the log-frequency
coordinate, for which the physical flux is defined as
$F_{g-1/2} = u_\nu E_{g-1/2}$. For the boundary conditions in the frequency
domain we assume zero radiation flux so that no radiation will leak at the
edges of the frequency domain. Equation (\ref{eq:closed}) can be discretized
with the one-stage second order upwind scheme
\begin{equation}
  E_g^{n+1} = E_g^* - \Delta t \frac{f_{g+1/2}-f_{g-1/2}}{\Delta(\ln \nu)},
\end{equation}
where time level $*$ is now the state after the hydro update and
the numerical flux is
\begin{eqnarray}
  f_{g-1/2} &=& u_\nu
  \left[E_g-\frac{1-{\rm C}}{2}\bar{\Delta}
    (E_{g+1}-E_{g},E_{g}-E_{g-1})\right], \qquad u_\nu\le0,\nonumber\\
  f_{g-1/2} &=& u_\nu
  \left[E_{g-1}+\frac{1-{\rm C}}{2}\bar{\Delta}
    (E_g-E_{g-1},E_{g-1}-E_{g-2})\right], \qquad u_\nu\ge0.\label{eq:mg9}
\end{eqnarray}
and we use the superbee limiter \citep{roe1986} for the limited slope
$\bar{\Delta}$.
The Courant-Friedrichs-Levi number ${\rm C} = |u_\nu|\Delta t/\Delta(\ln\nu)$
depends on the hydrodynamic time-step $\Delta t$. If C is larger than one,
the frequency advection is sub-cycled with the number of steps equal to the
smallest integer value larger than C.

\subsection{Implicit Diffusion and Energy Exchange}\label{sec:implicit}

The stiff parts of the radhyro equations are
solved implicitly in an operator split fashion. These stiff parts are the
radiation energy diffusion, electron heat conduction, and the energy exchange
between the electrons and each energy group $g$ and between the electrons and
ions. In this section, we will describe two implicit schemes that are
implemented: (1) solving for all radiation groups, electron and ion
temperatures in a coupled manner (Section \ref{sec:unsplit}), and
(2) solving each radiation group energy and the electron temperature
independently (Section \ref{sec:split}). Our strategy for resolution changes
is described in Appendix \ref{sec:reschange}, while the modifications for
the $rz$-geometry are explained in Appendix \ref{sec:rz}.

\subsubsection{Coupled Implicit Scheme}\label{sec:unsplit}

Discretizing the diffusion and energy exchange terms of equations
(\ref{eq:ionenergy})--(\ref{eq:electronenergy}), and (\ref{eq:groupenergy})
implicitly in time leads to
\begin{eqnarray}
  \frac{E_i^{n+1}-E_i^*}{\Delta t} &=& \sigma_{ie}^*(T_e^{n+1} - T_i^{n+1}),
  \label{eq:implionenergy}\\
  \frac{E_e^{n+1}-E_e^*}{\Delta t} &=& \sigma_{ie}^*(T_i^{n+1} - T_e^{n+1})
  + \nabla\cdot C_e^*\nabla T_e^{n+1}
  + \sum_{g=1}^G \sigma_g^*(E_g^{n+1} - B_g^{n+1}),
  \label{eq:implelectronenergy}\\
  \frac{E_g^{n+1}-E_g^*}{\Delta t} &=& \sigma_g^*(B_g^{n+1} - E_g^{n+1})
  + \nabla\cdot D_g^*\nabla E_g^{n+1},
  \label{eq:implgroupenergy}
\end{eqnarray}
where time level $*$ now corresponds to the state after the hydro update and
the frequency advection. The coupling coefficients $\sigma_{ie}^*$ and
$\sigma_g^*$ and the diffusion coefficients $C_e^*$ and $D_g^*$ are taken at
time level $*$ (frozen coefficients). One can either (1) solve the coupled
system of $G+2$ equations
(\ref{eq:implionenergy})--(\ref{eq:implgroupenergy}) implicitly or (2) solve
equation (\ref{eq:implionenergy}) for the ion internal energy $E_i^{n+1}$,
substitute the solution back into equation (\ref{eq:implelectronenergy}), and
solve the resulting reduced set of $G+1$ equations
(\ref{eq:implelectronenergy})--(\ref{eq:implgroupenergy})
implicitly. Here we describe the second scheme, because it is more
efficient, especially for small number of groups, e.g., for gray radiation
diffusion. Note that if we did include ion heat conduction in
(\ref{eq:implgroupenergy}), then we would have to solve the entire coupled
system of equations.

First we introduce the ion Planck function $B_i = aT_i^4$ as a new variable
similar to the electron Planck function $B = aT_e^4$, and replace $E_i$ and
$E_e$ with these variables using the chain rule
\begin{equation}
  \frac{\partial E_i}{\partial t} = \frac{\partial E_i}{\partial T_i}
  \frac{\partial T_i}{\partial B_i} \frac{\partial B_i}{\partial t}
  = \frac{C_{Vi}}{4aT_i^3}\frac{\partial B_i}{\partial t}, \qquad
  \frac{\partial E_e}{\partial t} = 
  \frac{C_{Ve}}{4aT_e^3}\frac{\partial B}{\partial t},\label{eq:freeze1}
\end{equation}
in which $C_{Vi}$ and $C_{Ve}$ are the specific heats of the ions and
electrons, respectively. Now equation (\ref{eq:implionenergy}) can be replaced
with
\begin{equation}
  B_i^{n+1} = B_i^* + \Delta t \sigma_{ie}' (B^{n+1} - B_i^{n+1}),
  \label{eq:planckion}
\end{equation}
where
\begin{equation}
  \sigma_{ie}' = \sigma_{ie}^*\frac{4aT_i^3}{C_{Vi}}
  \frac{1}{a(T_e+T_i)(T_e^2+T_i^2)}, \label{eq:freeze2}
\end{equation}
is again taken at time level $*$. The numerator comes from
$(T_e^4-T_i^4)/(T_e-T_i)$. Equation (\ref{eq:planckion}) can be solved
for $B_i^{n+1}$. This result can be substituted into the electron internal
energy equation (\ref{eq:implelectronenergy}) to obtain
\begin{equation}
  \frac{C_{Ve}'}{\Delta t} (B^{n+1} - B^*)
  = \sigma_{ie}''(B_i^* - B^{n+1})
  + \nabla\cdot C_e'\nabla B^{n+1}
  + \sum_{g=1}^G \sigma_g^*(E_g^{n+1} - w_g^* B^{n+1}),
\end{equation}
where we have introduced new coefficients at time level $*$:
\begin{equation}
  \sigma_{ie}'' = \frac{C_{Vi}}{4aT_i^3}
  \frac{\sigma_{ie}'}{1+\Delta t \sigma_{ie}'}, \qquad
  C_e' = \frac{C_e^*}{4aT_e^3}, \qquad
  C_{Ve}' = \frac{C_{Ve}^*}{4aT_e^3}. \label{eq:freeze3}
\end{equation}
The Planck weights $w_g^* = B_g^*/B^*$ satisfy $\sum_g w_g=1$.
It is convenient to introduce the changes $\Delta B = B^{n+1} - B^*$ and
$\Delta E_g = E_g^{n+1} - E_g^*$ to arrive at
\begin{eqnarray}
  \left[ \frac{C_{Ve}'}{\Delta t} + \sigma_{ie}''
    - \nabla\cdot C_e'\nabla \right] \Delta B
  &-& \sum_{g=1}^G \sigma_g^* (\Delta E_g - w_g^* \Delta B)
  = \sigma_{ie}''(B_i^* - B^*) \nonumber \\
  &+& \nabla\cdot C_e'\nabla B^* 
  + \sum_{g=1}^G\sigma_g^*(E_g^* - w_g^* B^*), \label{eq:implelectronnoncos}\\
  \left[ \frac{1}{\Delta t} - \nabla\cdot D_g^*\nabla \right] \Delta E_g
  &-& \sigma_g^*(w_g^*\Delta B - \Delta E_g) = \sigma_g^*(w_g^*B^* - E_g^*)
  + \nabla\cdot D_g^* \nabla E_g^*.\label{eq:implgroupnoncons}
\end{eqnarray}
This is a coupled system of $G+1$ linearized equations for the changes
$\Delta B$ and $\Delta E_g$. The right hand sides are all at time level *.

A discrete set of equations are obtained by applying the standard finite volume
method to the equations (\ref{eq:implelectronnoncos}) and
(\ref{eq:implgroupnoncons}) and partitioning the domain in a set of control
volumes $V_i$, enumerated by a single index $i=1,\ldots, I$. As an example, the
fluxes $F_{gij}$ associated with the radiation diffusion operator may be
obtained by approximating the gradient of the group energy density with a
simple central difference in the uniform part of the mesh:
\begin{equation}
  - \int_{V_i} \nabla\cdot(D_g\nabla E_g)dV
  = \sum_{j} F_{gij}
  = \sum_{j} S_{ij}D_{gij}\frac{E_{gi} - E_{gj}}{|{\bf x}_i - {\bf x}_j|},
  \label{eq:diffusionflux}
\end{equation}
where the index $j$ enumerates the control volumes which have a common face
with the control volume $i$, the face area being $S_{ij}$, and the distance
between the cell centers is $|{\bf x}_i - {\bf x}_j|$. Note that we assumed
here an orthogonal mesh. Generalization to curvilinear grids can be done
as shown in \citet{toth2008}. The diffusion
coefficients at the face are obtained by simple averaging of the cell centered
diffusion coefficient: $D_{gij} = (D_{gi}+D_{gj})/2$. The discretization
of the diffusion operator at resolution changes is described in Appendix
\ref{sec:reschange}.

The linear system (\ref{eq:implelectronnoncos})--(\ref{eq:implgroupnoncons})
can be written in a more compact form as the linearized implicit backward
Euler scheme
\begin{equation}
  \left( {\bf I} - \Delta t \frac{\partial {\bf R}}{\partial {\bf U}} \right)
  \Delta {\bf U} = \Delta t {\bf R}({\bf U}^*), \label{eq:euler}
\end{equation}
where ${\bf U}$ are the $I\times(G+1)$ state variables $B$ and $E_g$ for
all $I$ control volumes, and $\Delta {\bf U} = {\bf U}^{n+1} - {\bf U}^*$.
${\bf R}$ is defined by the spatially discretized
version of the right hand side of equations (\ref{eq:implelectronnoncos}) and
(\ref{eq:implgroupnoncons}). The matrix ${\bf A} = {\bf I}-\Delta t
\partial {\bf R}/ \partial {\bf U}$ is a $I\times I$ block matrix consisting of
$(G+1) \times (G+1)$ sub-matrices. This matrix ${\bf A}$ is in general
non-symmetric due to the Planck weight $w_g^*$ in the energy exchange between
the radiation and electrons. To solve this system we use Krylov sub-space type
iterative solvers, like GMRES \citep{saad1986} or Bi-CGSTAB
\citep{vandervorst1992}. To accelerate the convergence of
the iterative scheme, we use a preconditioner. In the current implementation
of CRASH, we use the Block Incomplete Lower-Upper decomposition (BILU)
preconditioner, which is applied for each adaptive mesh refinement block
independently. For gray radiation diffusion the Planck weight is one and
the matrix ${\bf A}$ can be proven to be symmetric positive definite (SPD) for
commonly used boundary conditions (see for example \citet{edwards1996}).
In that case we can use a preconditioned conjugate gradient (PCG) scheme
(see for instance \citet{eisenstat1981}).

For some verification tests, we can attempt to go second order in time under
the assumption of temporally constant coefficients using the Crank-Nicolson
scheme
\begin{equation}
  \frac{{\bf U}^{n+1} - {\bf U}^*}{\Delta t} = (1 - \alpha) {\bf R}({\bf U}^*)
  + \alpha {\bf R}({\bf U}^{n+1}),
\end{equation}
with $\alpha=1/2$. The implicit residual can again be
linearized ${\bf R}({\bf U}^{n+1}) = {\bf R}({\bf U}^*) +
(\partial {\bf R}/\partial {\bf U})^* \Delta{\bf U}$ to obtain the linear
system of equations
\begin{equation}
  \left( {\bf I} - \alpha \Delta t \frac{\partial {\bf R}}{\partial {\bf U}}
  \right) \Delta {\bf U} = \Delta t {\bf R}({\bf U}^*). \label{eq:crank}
\end{equation}
We use the same iterative solvers as for the backward Euler scheme.

Finally, we show how we use the solution $\Delta B$ and $\Delta E_g$ for
$g=1,\ldots, G$ from the non-conservative equations
(\ref{eq:implelectronnoncos})
and (\ref{eq:implgroupnoncons}) to advance the solution of the original
equations (\ref{eq:implionenergy})--(\ref{eq:implgroupenergy}) and still
conserve the total energy. One needs to express the fluxes and energies on
the right hand side in the latter equations in terms of $B^{n+1}$ and
$E_g^{n+1}$ while still keeping the coefficients frozen. After some algebra
we obtain
\begin{eqnarray}
  E_i^{n+1} &=& E_i^* + \Delta t \sigma_{ie}'' (B^{n+1} - B_i^*), \\
  E_e^{n+1} &=& E_e^* + C_{Ve}'(B^{n+1} - B^*), \\
  E_g^{n+1} &=& E_g^* + \Delta E_g.
\end{eqnarray}
This update conserves the total energy to round-off error. Note that at this
final stage, taking too large time step may result in negative ion internal
energy $E_i^{n+1}$ if $B^{n+1} \ll B_i^*$ and negative electron internal
energy $E_e^{n+1}$ if $B^{n+1} \ll B^*$. If this happens, the advance might be
redone with a smaller time step, to limit the drop in $B$, or by some other
timestep control scheme. A generalization of the conservative update to the
Crank-Nicolson scheme is also implemented for verification tests with time
constant coefficients.

For completeness, we mention that in the absence of radiation we solve during
the implicit step for the temperatures $T_e$ and $T_i$ instead of the
radiation energy-like variables $a T_e^4$ and $a T_i^4$. In that case the
corresponding matrix ${\bf A}$ is always SPD.
In principle, the formulation in temperatures can be generalized to
radiation as well. In \citet{landau1980}, a spectral temperature
$T_\nu(E_\nu,\nu)$ is defined, such that the spectral energy density is
locally equal to the spectral Planckian energy density at the temperature
$T_\nu$: $E_\nu = B_\nu(T_\nu,\nu)$. This relationship is a one-to-one map.
A group temperature, $T_g$, can also be introduced as the discrete analog such
that the group energy density can be obtained by
\begin{equation}
  E_g(T_g) = \int_{\nu_{g-1/2}}^{\nu_{g+1/2}} B_\nu (T_g,\nu)d\nu.
\end{equation}
The equation (\ref{eq:implgroupenergy}) can be recast as equation for
the group temperature $T_g$. This introduces the group specific heat of the
radiation $C_g = dE_g/dT_g$. The set of equations
(\ref{eq:implionenergy})--(\ref{eq:implgroupenergy}) reformulated as an
implicit backward Euler scheme for the temperatures $T_i$, $T_e$, and $T_g$
can in a similar way as in \citet{edwards1996}
be proven to be SPD. While this scheme has the advantage of being SPD,
the conservative update of the group energy density
$E_g^{n+1} = E_g^* + C_g^* \Delta T_g$ might result in negative energy
density $E_g^{n+1}$ for too large time steps.

\subsubsection{Decoupled Implicit Scheme}\label{sec:split}

The coupled implicit scheme of Section \ref{sec:unsplit} requires the solution
of a large system of equations ($G+1$ variables per mesh cell). The
preconditioning for such a system can be computationally expensive
and requires overall lots of memory. We therefore also implemented a
decoupled implicit scheme that solves each equation independently.

For some applications, the electron temperature does not change much due to
energy exchange with the radiation. This is typically so if the electrons have
a much larger energy density than the radiation, so that $T_e$ changes little
due to the interaction with the radiation in a single time step. In that case,
we solve first for the electron and ion temperatures without the contributions
from the radiation-electron energy exchange. Let again time level $*$ indicate
the state after the hydro update and frequency advection, and freeze again
$C_e^*$, $D_g^*$, $\sigma_{ie}^*$, $\sigma_g^*$ at time level $*$.
Discretization in time now leads to
\begin{eqnarray}
  \frac{E_i^{n+1}-E_i^*}{\Delta t} &=& \sigma_{ie}^*(T_e^{**} - T_i^{n+1}),
  \label{eq:splitionenergy}\\
  \frac{E_e^{**}-E_e^*}{\Delta t} &=& \sigma_{ie}^*(T_i^{n+1} - T_e^{**})
  + \nabla\cdot C_e^*\nabla T_e^{**},
  \label{eq:splitelectronenergy}
\end{eqnarray}
where the time level $**$ of $E_e$ indicates that we still have to do an
extra update to time level $n+1$ with the radiation-electron energy exchange.
Each radiation group energy density is solved independently using time level
$*$ for the electron temperature in $B_g^*$:
\begin{equation}
  \frac{E_g^{n+1}-E_g^*}{\Delta t} = \sigma_g^*(B_g^* - E_g^{n+1})
  + \nabla\cdot D_g^*\nabla E_g^{n+1},
  \label{eq:splitgroupenergy}
\end{equation}
where we have exploited the assumption that $B_g^*$ is not stiff.

Equations (\ref{eq:splitionenergy})--(\ref{eq:splitgroupenergy}) can be recast
in equations for the $G+1$ independent changes  $\Delta B = B^{**} - B^*$ and
$\Delta E_g = E_g^{n+1}-E_g^*$:
\begin{eqnarray}
  \left[ \frac{C_{Ve}'}{\Delta t} + \sigma_{ie}''
    - \nabla\cdot C_e'\nabla \right] \Delta B
  &=& \sigma_{ie}''(B_i^* - B^*) + \nabla\cdot C_e'\nabla B^*,
  \label{eq:splitelectronnoncos} \\
  \left[ \frac{1}{\Delta t} +\sigma_g^* - \nabla\cdot D_g^*\nabla \right]
  \Delta E_g &=& \sigma_g^*(w_g^*B^* - E_g^*)
  + \nabla\cdot D_g^* \nabla E_g^*.\label{eq:splitgroupnoncons}
\end{eqnarray}
where we have used the definitions (\ref{eq:freeze1}), (\ref{eq:freeze2}), and
(\ref{eq:freeze3}) of the coefficients, frozen at time level $*$.
Each equation for the changes is in the form of the linearized implicit
backward Euler scheme (\ref{eq:euler}) and can be solved independently with
iterative solvers like GMRES and Bi-CGSTAB using a BILU preconditioner. As
long as the boundary conditions are such that the matrices are symmetric and
positive definite, the preconditioned conjugate gradient method might also be
used.

In a similar manner as with the coupled implicit scheme, a conservative update
for the energy densities can be derived as
\begin{eqnarray}
  E_g^{n+1} &=& E_g^* + \Delta E_g, \\
  E_i^{n+1} &=& E_i^* + \Delta t \sigma_{ie}'' (B^{**} - B_i^*), \\
  E_e^{n+1} &=& E_e^* + C_{Ve}'(B^{**} - B^*)
  + \Delta t \sum_{g=1}^G \sigma_g^* (E_g^{n+1}-w_g^*B^*),
  \label{eq:conseeupdate}
\end{eqnarray}
that preserve the total energy to roud-off errors. The main difference
between the conservative update in the coupled and decoupled
schemes is that here the energy exchange between the radiation and electrons is
added afterwards as the last term in equation (\ref{eq:conseeupdate}).

This scheme requires less computational time for preconditioning and the
Krylov solver than the coupled implicit algorithm, however it generally needs
more message passing in parallel computations. It is therefore not always
guaranteed that the decoupled scheme is faster. The memory usage is always
smaller.

\subsection{Boundary Conditions}\label{sec:boundary}

The CRASH code allows for any user specified type of boundary conditions.
Several commonly used boundary conditions are readily available in the main
code for convenience, e.g., fixed, extrapolation with zero gradient, periodic,
and reflective boundary conditions.

For the radiation field, we have implemented a zero or fixed incoming flux
boundary condition that is used instead of the extrapolation with zero
gradient. This type of boundary condition is useful if there are no sources of
radiation outside the computational domain and we assume that outflowing
radiation does not return back into the computational domain (zero albedo).
Note that simple extrapolation with zero gradient can make the radiation
diffusion problem ill-posed. The boundary condition is derived as follows:
Radiation diffusion approximation corresponds to a linear-in-angle intensity
distribution
\begin{equation}
  I_g = \frac{c}{4\pi}E_g + \frac{3}{4\pi} {\bf F}_g\cdot{\bf n},
  \label{eq:closure}
\end{equation}
so we can calculate the radiation flux
through a boundary surface. If we define the outward pointing normal vector of
the boundary as ${\bf n}_b$, the net flux of radiation energy inward through
this boundary is 
\begin{equation}
  F^{\rm in}_g
  = -\int_{{\bf n}\cdot{\bf n}_b<0} {\bf n}_b\cdot{\bf n} I_gd\Omega
  = \frac{cE_g}{4} -\frac{1}{2}{\bf n}_b\cdot{\bf F}_g,
\end{equation}
where the closure (\ref{eq:closure}) is used. In the radiation diffusion model,
the flux is written as ${\bf F}_g = -D_g\nabla E_g$, where the diffusion
coefficient $D_g$ is a nonlinear function of $E_g$ and $\nabla E_g$ in a flux
limited diffusion model. The boundary condition satisfies
\begin{equation}
  E_g + \frac{2D_g}{c} {\bf n}_b\cdot\nabla E_g = \frac{4}{c}F_g^{in}.
\label{eq:fixbc}
\end{equation}
For the left boundary in the $x$-direction, for instance, this can be
discretized as
\begin{equation}
  \frac{E_{g0}+E_{g1}}{2} - \frac{2D_g}{c} \frac{E_{g1} - E_{g0}}{\Delta x}
  = \frac{4}{c}F_g^{in},
\end{equation}
where the index 1 corresponds to the last physical cell and 0 to the ghost
cell. This equation can be solved for the ghost cell value. For zero
incoming radiation flux boundary conditions we set $F_g^{in}=0$.

\section{CODE VERIFICATION}\label{sec:verification}

To test the CRASH as well as the BATS-R-US and SWMF codes,
we have implemented numerous tests. These tests are subdivided in two
categories: functionality tests and verification tests. Both test suites are
performed automatically and return pass or fail messages depending on whether
or not certain predefined tolerance criteria are met. This automated testing
process provides a software quality confidence especially when used in
combination with a software version control system like CVS (Concurrent
Versions System) to recover previous correctly performing code.

The functionality tests are performed nightly on several computer platforms
with different compilers and number of processors. They consist of unit tests
and full system tests. The unit tests are designed to test a particular unit,
for example a linear equation solver. The full system tests on the other
hand, exercise the code in the way the end-users will use it for their
research applications. We always try to cover as much code as possible with
these tests so that we can discover bugs and other unwanted side effects
early on.

To test the correctness of the implemented algorithms we have also constructed
a suite of verification tests. This suite is executed daily on a dedicated
parallel computer and runs specific simulations to quantify against analytic
and semi-analytic solutions, whenever possible. The CRASH test repository
currently covers a wide range of tests for hydrodynamics, multi-material
advection methods, gray and multigroup radiation diffusion, heat conduction,
to mention a few. These are performed to test for grid and/or time convergence,
as deemed necessary. We also simulate full system laboratory experiment
configurations in various geometries, dimensionality, and
physics fidelity. The results are either validated against laboratory
experiments or simply used to check that the code keeps performing these
simulations as expected. Once a week, we also perform a parallel
scalability test on a large parallel computer to verify that the code does not
degrade in performance during further development of the software.

In the following sub-sections, we highlight some specific verification
tests related to the implicit radiation (Section \ref{sec:radverify}) and
heat conduction (Section \ref{sec:condverify}) solver. The tests cover both
Cartesian and $rz$-geometry, and some of them also involve the hydrodynamic
solver. We demonstrate a 3D full system test in Section \ref{sec:fullsystem}
and describe the parallel scalability in Section \ref{sec:parallel}.

\subsection{Error Assessment}

For the assessment of the accuracy of the solutions in the test suites, an
appropriate definition of the numerical errors have to be defined. We will use
two types of errors to quantify the verification analysis:
The relative L1 error is defined as 
\begin{equation}
  E_{\mbox{L1}} = \sum_{\alpha=1}^N \frac{\sum_{i=1}^I |{\bf U}_{\alpha i}
    -{\bf V}_{\alpha i}|}{\sum_{i=1}^I|{\bf V}_{\alpha i}|},
\end{equation}
where $\alpha=1,\ldots, N$ indexes the state variables of numerical solution
vector ${\bf U}$ and the reference solution ${\bf V}$, and $i=1,\ldots, I$
indexes the grid cells of the entire computational domain. For test problems
with smooth solutions, we will also use the relative maximum error defined by
\begin{equation}
  E_{\mbox{L$\infty$}} = \sum_{\alpha=1}^N \frac{\max_{i=1,\ldots,I}
    |{\bf U}_{\alpha i}-{\bf V}_{\alpha i}|}{\max_{i=1,\ldots,I}|
    {\bf V}_{\alpha i}|}.
\end{equation}
Quite often the reference solution is defined on a grid with higher resolution
than that of the numerical solution. In that case, we first coarsen the
reference solution to the resolution of the numerical solution. 

\subsection{Radiation Tests}\label{sec:radverify}

\subsubsection{Su-Olson Test}

\citet{suolson1996} developed a one-dimensional Marshak wave test, to check
the accuracy of the scheme and the correctness of the implementation of the
time-dependent non-equilibrium gray radiation diffusion model. In this test,
radiation propagates through a cold medium that is initially in
absence of radiation. The equations are linearized by the choice of the
specific heat of the material $C_V = 4aT^3$ as well as by setting the Rosseland
and Planck opacities to the same uniform and time-independent constant
$\kappa_R=\kappa_P=\kappa$. The cold medium is defined on a half-space of the
slab geometry $0\leq x<\infty$. At the boundary on the left, a radiative
source is specified, creating an incident radiation flux of $F^{in}=aT_{in}^4$,
where $T_{in}=1\,$keV. As time progresses, the radiation diffuses
through the initially cold medium and by energy exchange between radiation and
matter, the material temperature will rise. In \citet{suolson1996}, a
semi-analytical solution is derived for the time evolution of the radiation
energy and material temperature. We will use this solution for this
verification test.

\begin{figure}
\plotone{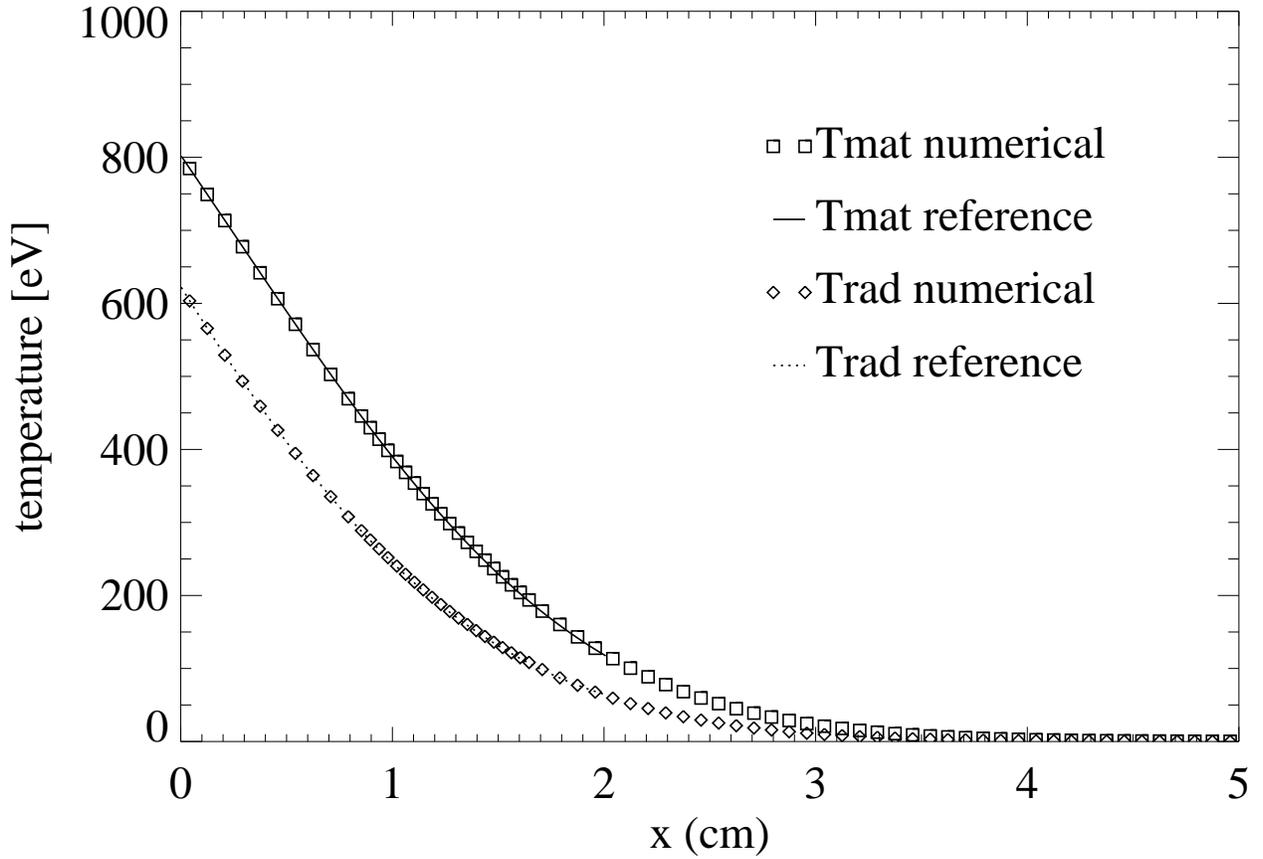}
\caption{The material ($T_{\mbox{mat}}$) and radiation ($T_{\mbox{rad}}$)
temperature solution of the
\cite{suolson1996} non-equilibrium Marshak radiation diffusion problem
obtained with the CRASH code on a non-uniform grid. The reference temperatures
of the analytical method of \cite{suolson1996} are shown as lines.}
\label{fig:suolsontemp}
\end{figure}

For convenience, we locate the right boundary at the finite distance $x=5\;$cm
and impose a zero incoming radiation flux on that boundary. We decompose the
computational domain in 6 grid blocks at the base level with 10 cells per
block. Between $x=5/6\;$cm and $x=5/3\;$cm, the domain is refined by one
level of AMR. During the time evolution, the radiation diffuses to the right
through the resolution changes. The system is time evolved with the implicit
radiation diffusion solver by using a preconditioned conjugate gradient method
till the final time $0.02\;$ns.  The solver steps through a series of fixed
time steps of $5\times10^{-4}\;$ns and we use a Crank-Nicolson approach to
achieve second order accurate time-integration. Note that this is possible
because coefficients of the matrix to be solved are not time dependent. The
computed radiation and material temperatures at the final time are shown in
Figure \ref{fig:suolsontemp} and agree well with the semi-analytical solution.

\begin{figure}
\plotone{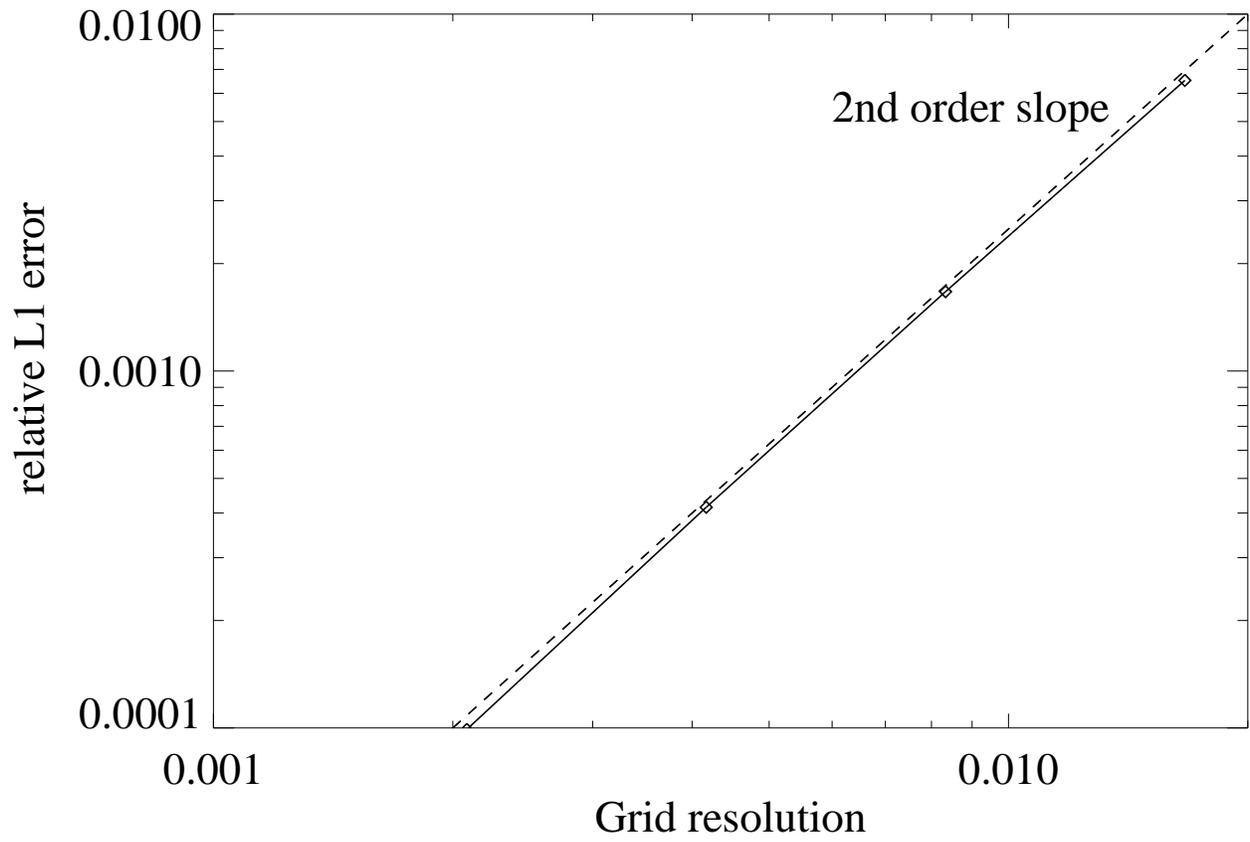}
\caption{The relative L1 error for the Su-Olson test on a non-uniform grid.}
\label{fig:suolsonerror}
\end{figure}

Figure \ref{fig:suolsonerror} shows the relative L1 error of the radiation
and material teperatures versus increasing grid resolution of the base level
grid. We did not use the semi-analytical solution as the reference, since it is
difficult to get an accurate enough solution with the quadrature method as
mentioned by \citet{suolson1996}. Instead, we use a very high
resolution (1920 cells) numerical reference solution obtained with the CRASH
code. Four different base level resolutions with 60, 120, 240, and 480 cells
are used to demonstrate the second order convergence. The time step is
proportional to the cell size $\Delta x$.

\subsubsection{Lowrie's Non-equilibrium Radhydro Solutions}
\label{sec:lowrie1}

\citet{lowrie2008} designed several shock tube problems for the
non-equilibrium gray radiation diffusion coupled to the hydrodynamic equations
that can be used for code verification. These solutions are planar radiative
shock waves where the material and radiation temperatures are out of
equilibrium near the shock, but far from the shock the flow is assumed to be
in radiative equilibrium. Depending on the Mach number of the pre-shock state,
a wide range of shock behavior can occur.
For the CRASH test suite, we selected a few of the semi-analytic solutions
from \citet{lowrie2008}. In this section we will describe the Mach 1.05 flow
with uniform opacities as an example. Here the shock is smoothed out by the
energy exchange with the diffusive radiation. Another more challenging Mach 5
problem with non-uniform opacities will be described in Section
\ref{sec:lowrie3}.

\begin{figure}
\plotone{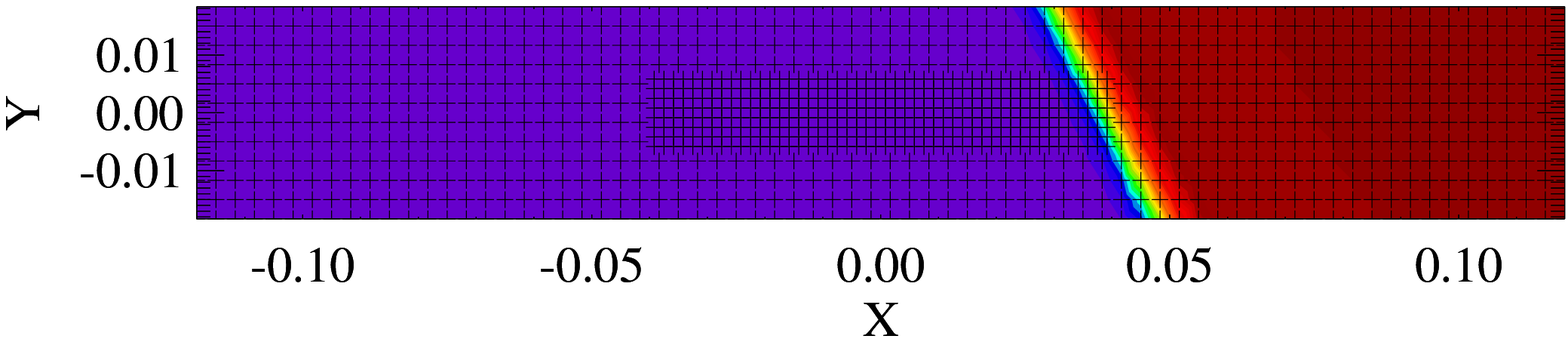}
\plotone{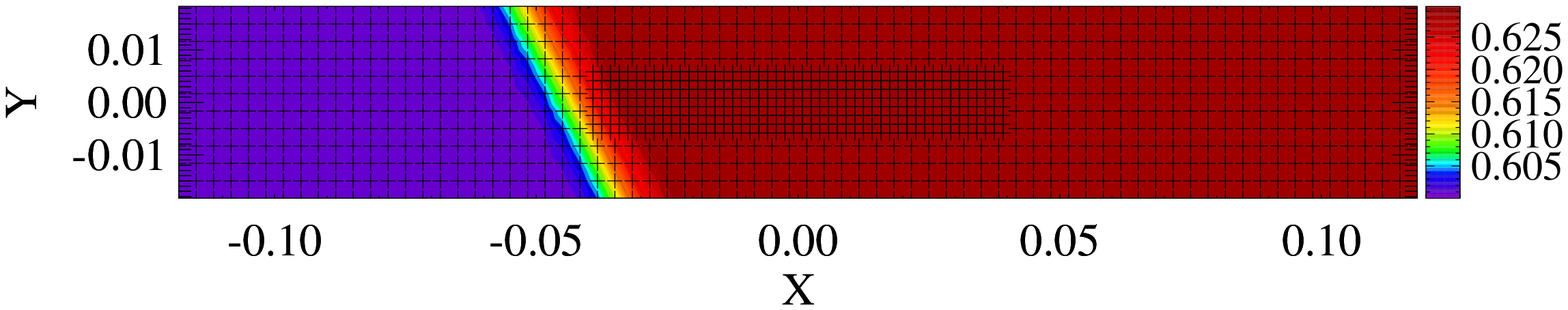}
\caption{Rotated shock tube test on a 2D AMR grid based on the Mach number
1.05 non-equilibrium gray radiation hydrodynamic test in \cite{lowrie2008}.
Shown is the radiation temperature in color contour at the initial (top panel)
and final (bottom panel) times. The black crosses indicate the cell centers.
}
\label{fig:lowrie1grid}
\end{figure}

The Mach 1.05 test is performed on a 2D non-uniform grid. The initial
condition is taken to be the same as the original steady state reference
solution. Since the system of equations is Galilean invariant, we can add an
additional velocity -1.05 so that the velocity on the left boundary is zero
while the smoothed shock will now move to the left. This new initial condition
as well as the velocity vector are rotated by
$\tan^{-1}(1/2)\approx 26.56^\circ$. This means that
there is a translational symmetry in the $(-1,2)$ direction of the $xy$-plane
as shown in Figure \ref{fig:lowrie1grid}. The computational domain is
$-0.12 < x < 0.12$ by $-0.02<y<0.02$ decomposed in $3\times 3$ grid blocks of
$24\times 4$ cells each. We apply one level of refinement
inside the region $-0.04 < x < 0.04$ by $-0.02/3<y<0.02/3$. The initial
smoothed shock starts at the right boundary of the refined grid and we time
evolve the solution till it reaches the resolution change on the left as
shown in Figure \ref{fig:lowrie1grid}. For the boundary conditions in the $x$
direction we use zero radiation influx conditions for the radiation field,
while zero gradient is applied to the remaining state variables. On the
$y$ boundaries, we apply a sheared zero gradient in the $(-1,2)$ direction
for all variables.

The hydrodynamic equations are time evolved with the HLLE scheme with a CFL
number 0.8. We use the generalized Koren limiter with $\beta=3/2$ for the
slope reconstruction. For the implicit radiation diffusion solver, we use the
GMRES iterative solver in combination with a BILU preconditioner.
The specific heat is time dependent since it depends on the density,
therefore the implicit scheme is only first order accurate in time.
To enable second order grid convergence for this smooth test problem, we
compensate this by reducing CFL number proportional to the grid cell size,
in other words $\Delta t\propto\Delta x^2$, so that second order accuracy
with respect to $\Delta x$ can be achieved. We increase the spatial resolution
by each time doubling the number of grid blocks at the base level in both the
$x$ and $y$ directions.

\begin{figure}
\begin{center}
{\resizebox{0.48\textwidth}{!}{\includegraphics[clip=]{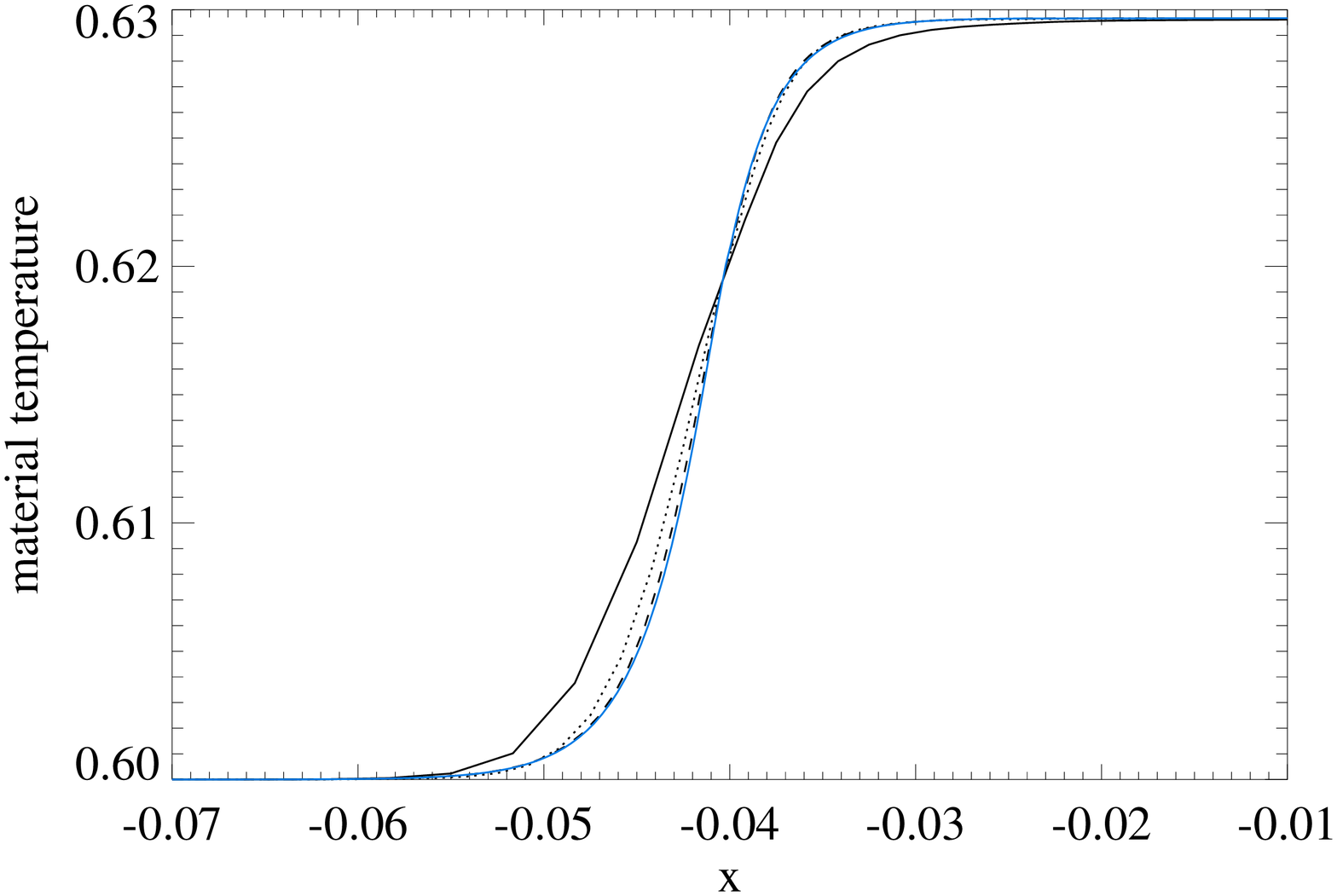}}}
{\resizebox{0.48\textwidth}{!}{\includegraphics[clip=]{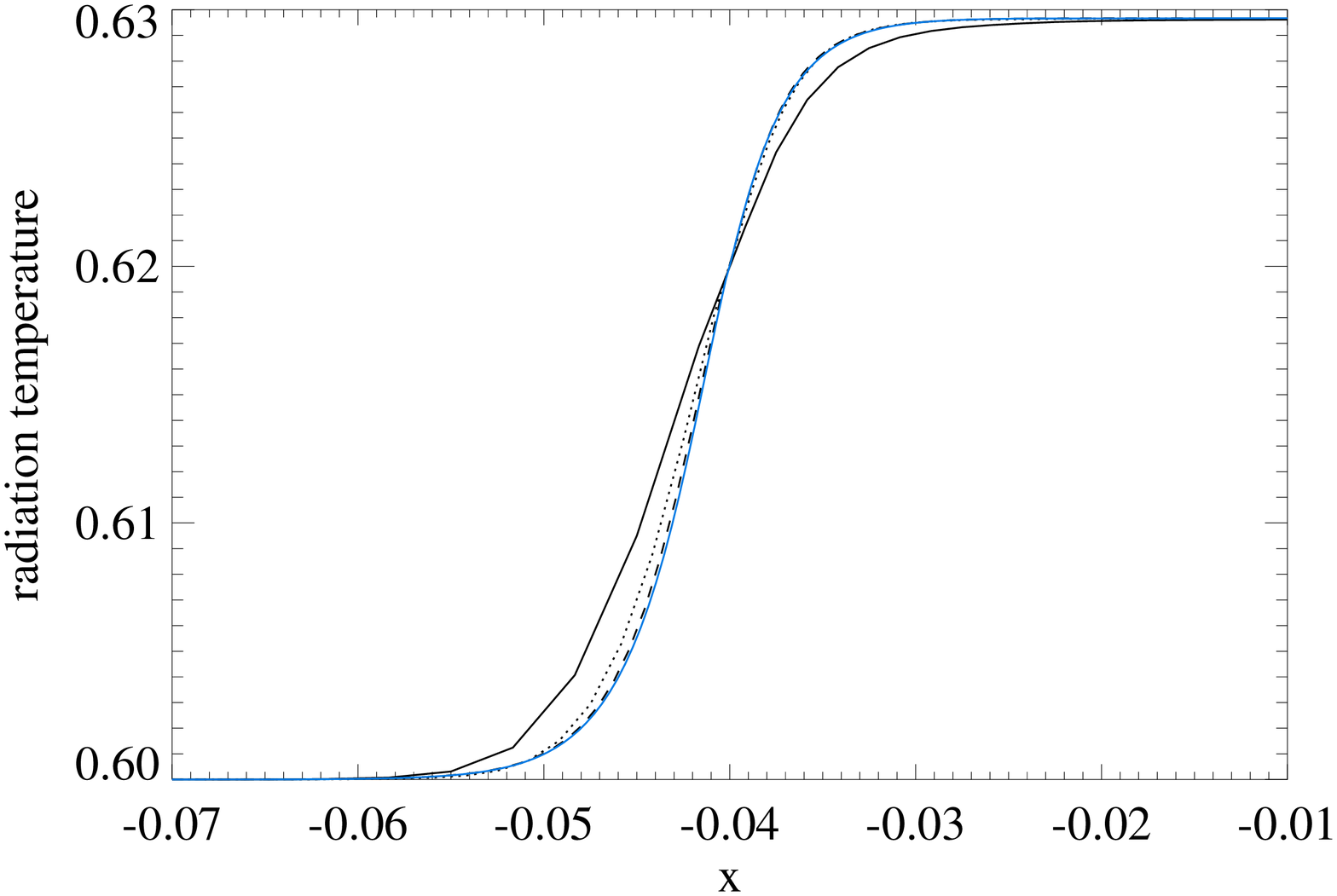}}}
\end{center}
\caption{The material (left panel) and radiation (right panel) temperatures for
the Mach 1.05 radiative shock tube problem at
the final time are shown in the $x$-direction. The solid, dotted, and dashed
lines correspond to three different grid resolutions, respectively. The blue
line is the semi-analytical reference solution of \cite{lowrie2008}.}
\label{fig:lowrie1temp}
\end{figure}

The convergence of the numerically obtained material and radiation
temperatures along the $y=0$ cut at the final time $t=0.07$ is shown in
Figure \ref{fig:lowrie1temp}. The solid, dotted, and dashed lines correspond to
the solutions with the $3\times 3$, $6\times 6$, and $12\times 12$ base level
grid blocks, respectively. The advected semi-analytical reference solution is
shown as a blue line for comparison.

\begin{figure}
\plotone{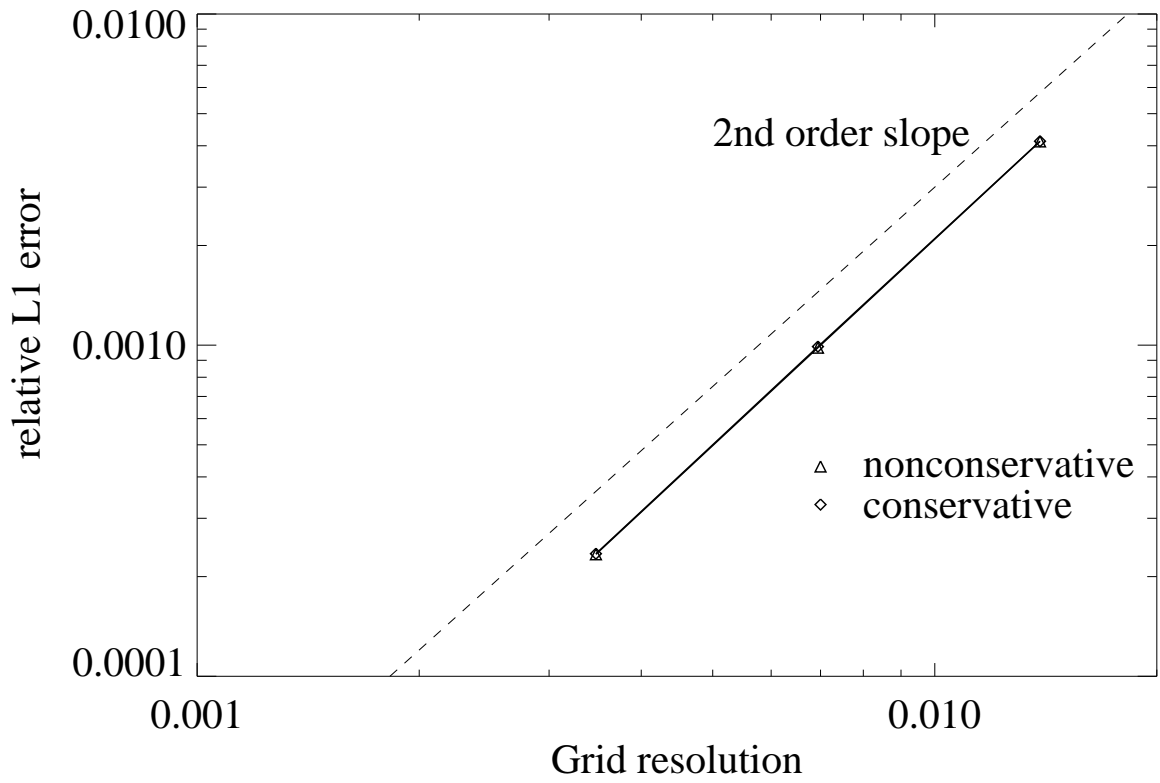}
\caption{The relative L1 error for the Mach 1.05 non-equilibrium radiation
diffusion test on a non-uniform grid. Both the non-conservative as well as
the conservative hydrodynamic schemes are tested.}
\label{fig:lowrie1error}
\end{figure}

To assess the order of accuracy, the grid convergence is shown in Figure
\ref{fig:lowrie1error} for the three resolutions. The relative L1 error
is calculated using the density, velocity components, and both the material
and radiation temperatures. We obtain second order convergence for both
the conservative as well as the non-conservative (using the pressure equation
instead of the total energy) hydrodynamic schemes. The latter scheme can be
used because in the Mach 1.05 test the hydro shock is smoothed out by the
interaction with the radiation.

\subsubsection{Double Light Front}

As a test for the multigroup radiation diffusion model we developed a
double light front test problem. This test is used to verify the
implementation of both the group diffusion
and flux limiters. At the light front the discontinuity in the radiation
field switches on the flux limiter. This limiter is used to correct the
radiation propagation speed in the optically thin free streaming regime. With
the light front test we can then check that we obtain the speed of light
propagation of the front and that the front maintains as much as possible
the initial discontinuity.

This test is constructed as follows: We use a 1D
computational domain of the size of $1\;$m in the $x$-direction. On this
domain, we initialize the two radiation group energy densities $E_g$ $(g=1,2)$
with a very small, positive number to avoid division by zero
in the flux limiter. Also the Rosseland mean opacities are set to a small
number corresponding to strong radiation diffusion, while the Planck mean
opacities are set to zero corresponding to an
optically thin medium. The radiation energy density of the first group enters
from the left boundary by applying a fixed boundary condition with value one in
arbitrary units. On the right boundary this group is extrapolated with
zero gradient. Note that these are the proper boundary conditions in the
free-streaming limit and not the diffusive flux boundary conditions described
in Section \ref{sec:boundary}. The second radiation group enters from the
right boundary with density one, and it is extrapolated with zero gradient
at the left boundary. We time evolve both groups for
$0.5\;$m$/c$ seconds. The analytic solution will then be two discontinuities
that have reached $x=0.5\;$m, since both fronts propagate with the speed of
light $c$.

\begin{figure}
\begin{center}
{\resizebox{0.48\textwidth}{!}{\includegraphics[clip=]{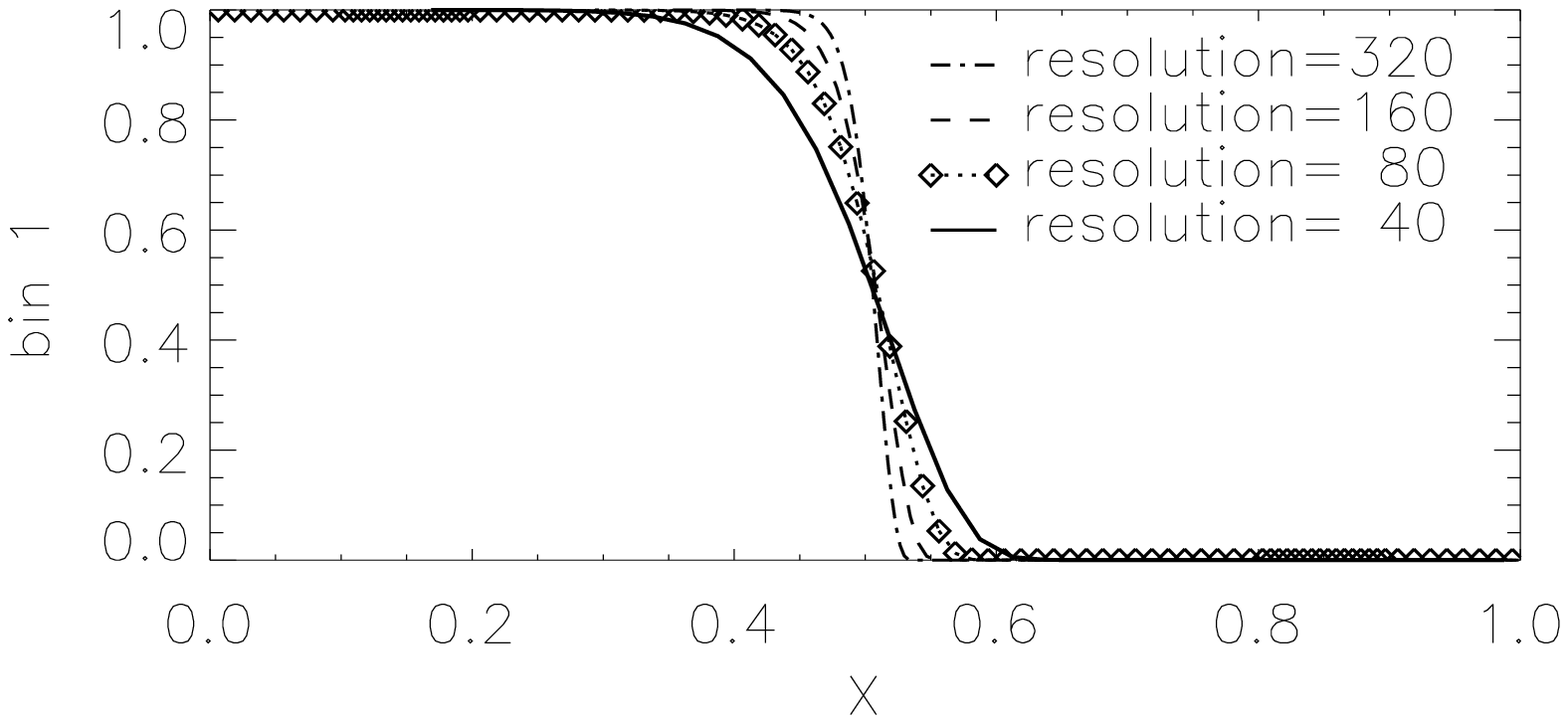}}}
{\resizebox{0.48\textwidth}{!}{\includegraphics[clip=]{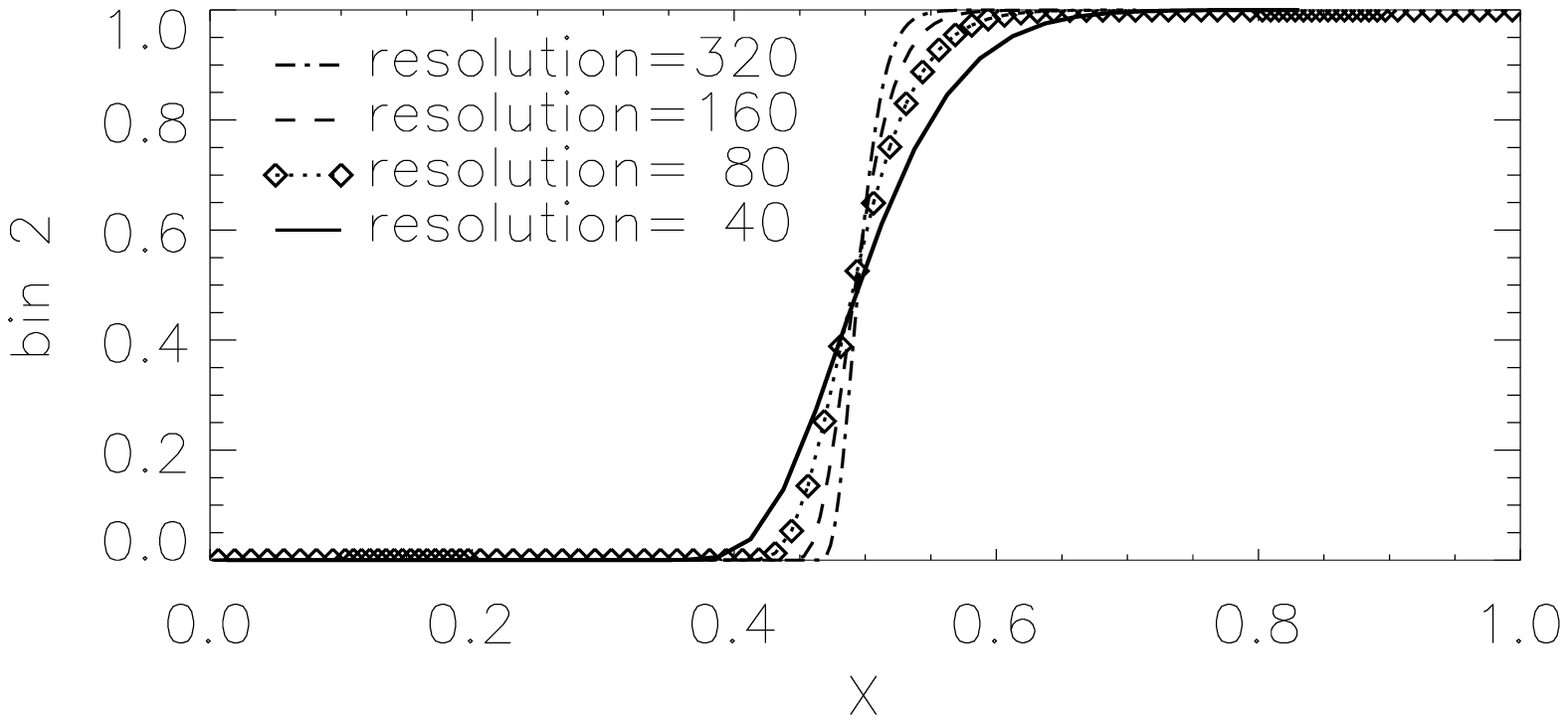}}}
\end{center}
\caption{Solutions for the 1D double light front test for 4 different
non-uniform grid resolutions. The radiation energy for group 1 (left panel)
enters from the left boundary, for group 2 (right panel) it enters from the
right boundary. The symbols for base resolution 80 shows one level of grid
refinement for $0.1<x<0.2$ and $0.8<x<0.9$.}
\label{fig:lightfront}
\end{figure}

The computational domain is non-uniform. In the coarsest resolution there are
10 grid blocks of
4 cells each at the base level. Inside the regions $0.1<x<0.2$ and $0.8<x<0.9$,
we use one level of refinement. The total time evolution is divided into
400 fixed time steps. We use GMRES for the radiation diffusion solver in
combination with a BILU preconditioner. For the grid convergence we reduce
the fixed time step quadratically with the grid resolution. This time step
reduction mimics second order discretization in time. In Figure
\ref{fig:lightfront}, the two group energy densities are shown for the
base level grid resolutions 40, 80, 160, and 320. Clearly, with increasing
number of cells, the solution converges towards the reference discontinuous
fronts at $x=0.5$.

\begin{figure}
\plotone{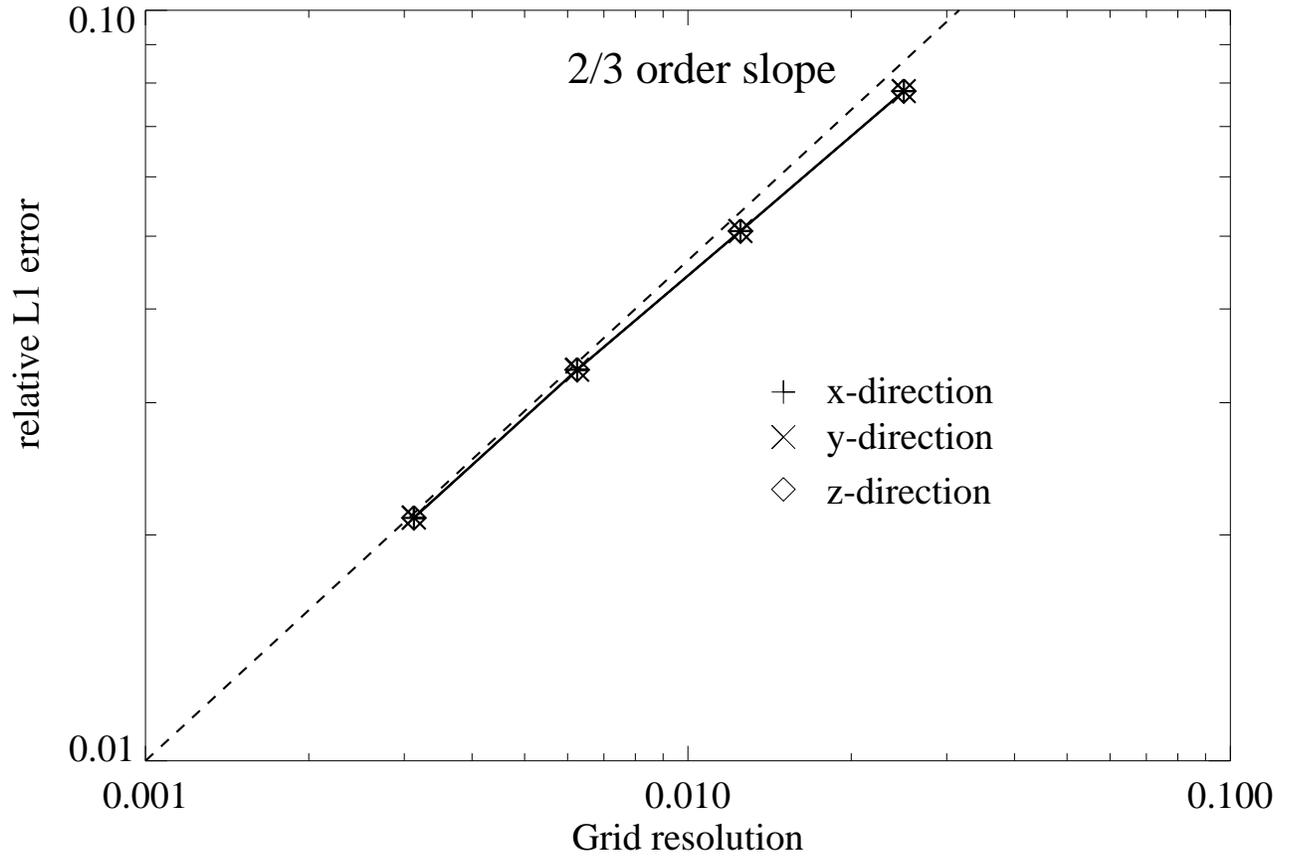}
\caption{Relative L1 error for the double light front test on a non-uniform
grid. The test is performed for the $x$, $y$, and $z$ directions.}
\label{fig:lightfronterror}
\end{figure}

In Figure \ref{fig:lightfronterror}, the grid convergence is shown for the
four resolutions. The relative L1 error is calculated using both radiation
group energy densities and compared to the analytical reference solution with
the discontinuities at $x=0.5$. In \citet{gittings2008}, it was stated
that for a second order difference scheme the convergence rate for a contact
discontinuity is 2/3. Indeed, we find this
type of convergence rate, due to the numerical diffusion of the
discontinuities, for the light front test. We have also performed the
tests in the $y$ and $z$ directions to further verify the implementation.

\subsubsection{Relaxation of Radiation Energy Test}

\begin{figure}
\plottwo{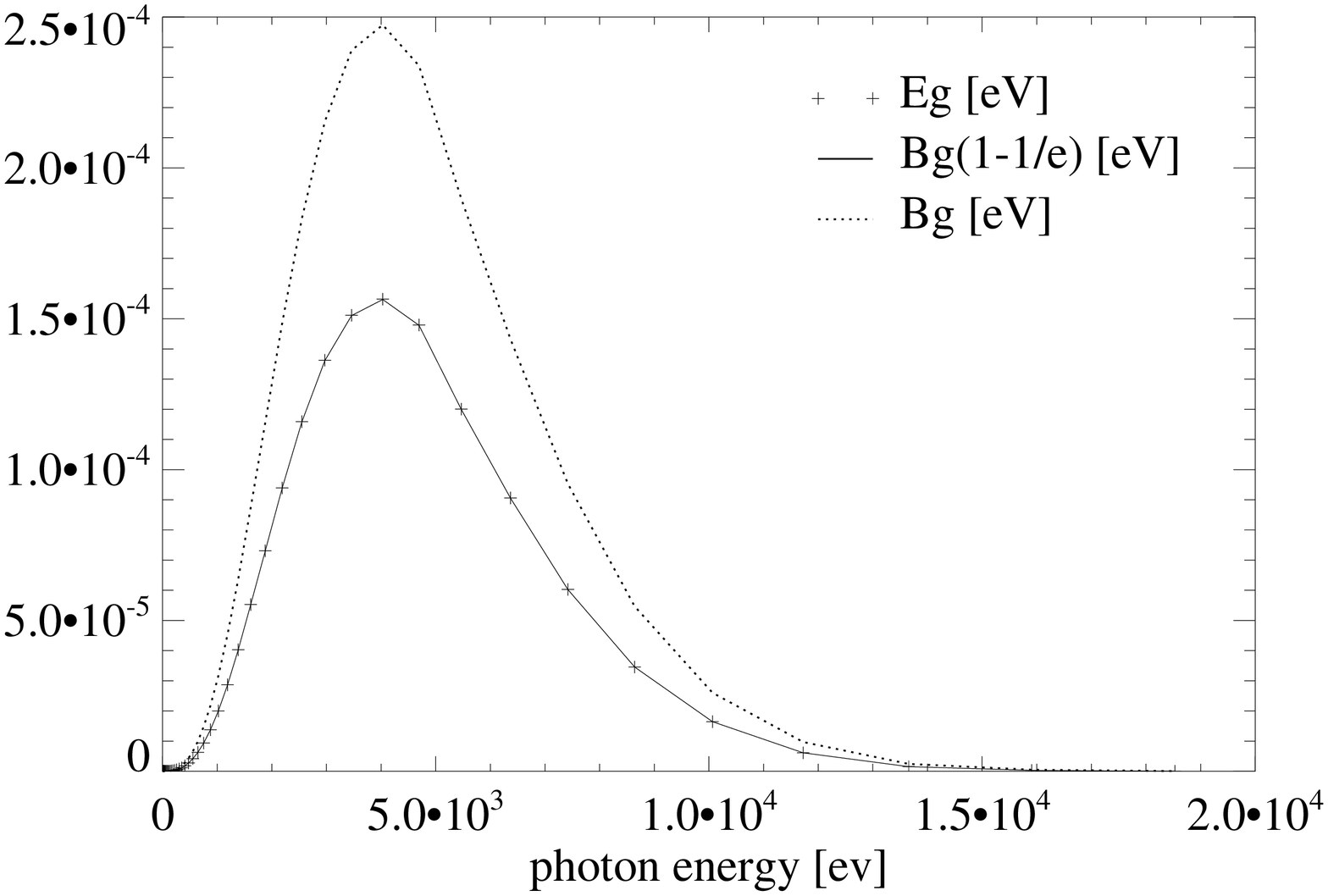}{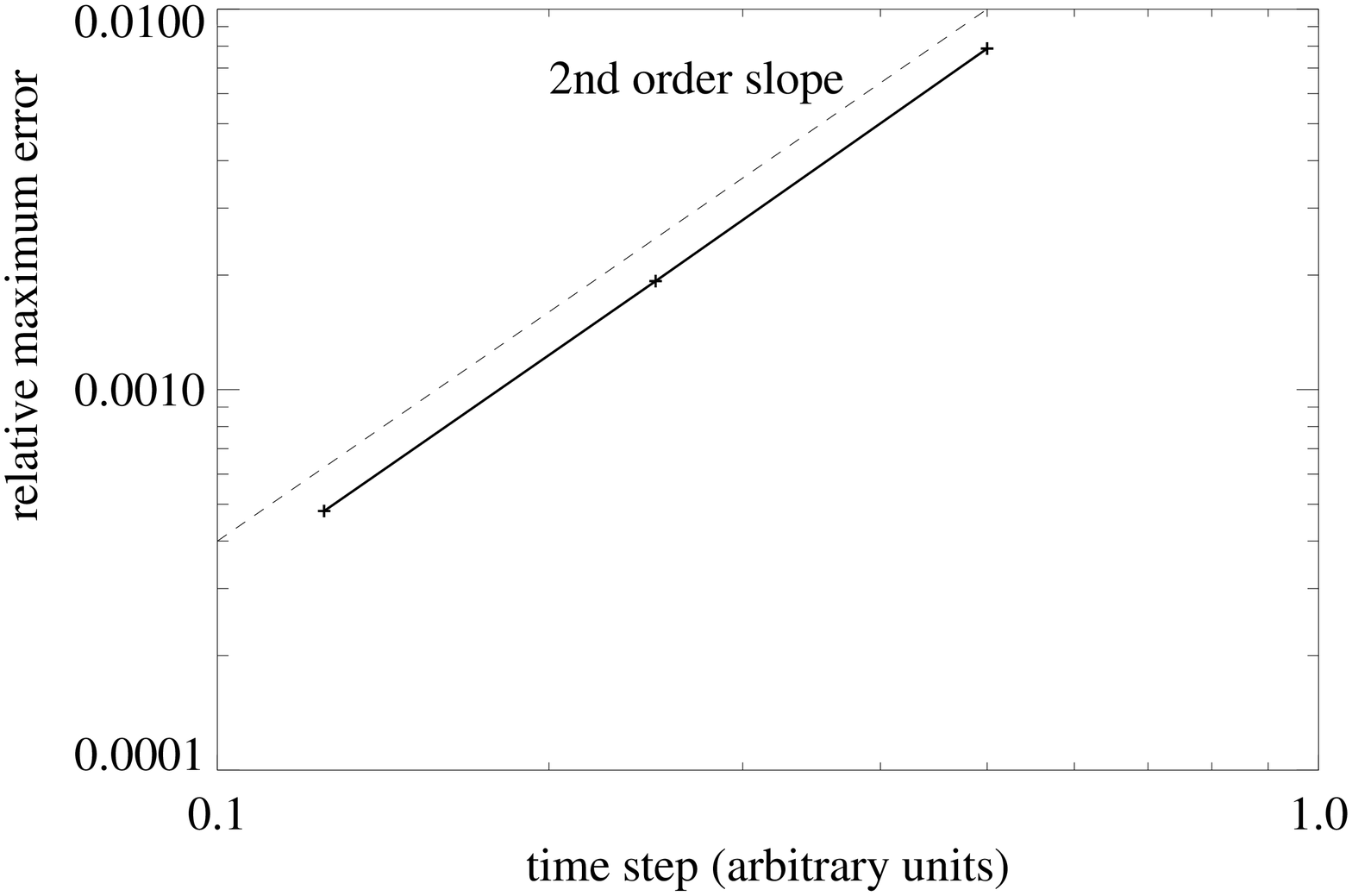}
\caption{The relaxation of radiation energy test for 80 groups. Left panel is
for the time independent spectrum $B_g$ (dotted line) and the group radiation
energy solution $E_g$ at time $1/\sigma_g$ (+ points) versus the photon
energies after 80 time steps. The
analytical reference solution is shown as a solid line. Right panel shows
the relative maximum error for $20$, $40$, and $80$ time steps
demonstrating second order convergence rate.}
\label{fig:infinitemedium}
\end{figure}

This test is designed to check the relaxation rate between the material and
the radiation. The energy exchange between the material and radiation groups
can be written as
\begin{eqnarray}
  C_V \frac{\partial T}{\partial t} &=& \sum_{g=1}^G \sigma_g(E_g - B_g), \\
  \frac{\partial E_g}{\partial t} &=& \sigma_g(B_g - E_g).
\end{eqnarray}
For a single radiation group, an analytic expression can be found to describe
the relaxation in time. However, for arbitrary number of groups,
a time dependent
analytic solution is less obvious, except for some rather artificial cases.
Here we make the assumption of an extremely large value of the specific heat
$C_V$ to make the analysis more tractable. In that case, the material
temperature is time independent, so that $B_g$ is likewise time independent.
The solution is then $E_g = B_g(1-e^{-\sigma_g t})$ assuming $E_g(t=0)=0$
initially. At time
$t=1/\sigma_g$, the group radiation energy density is $E_g = B_g(1-1/e)$.
Note that this test only needs one computational mesh cell in the spatial
domain. We set $T=1$\,keV and the resulting Planckian spectrum, defined by
$B_g$, is depicted by the dotted line in the left panel of Figure
\ref{fig:infinitemedium}. We use 80 groups logarithmically distributed over
the photon energy domain in the range of 0.1\,eV to 20\,keV. The computed
$E_g$ at time $t=1/\sigma_g$ are shown as + points. For the simulation we used
the GMRES iterative solver and the Crank-Nicolson scheme.
To assess the error, we repeated the test with time steps of $1/20$,
$1/40$, and $1/80$ of the simulation time. The second order convergence rate is
demonstrated in the right panel of Figure \ref{fig:infinitemedium}.

\subsection{Heat Conduction Tests}\label{sec:condverify}

\subsubsection{Uniform Heat Conduction in rz-geometry}

This test is designed to verify the implicit heat conduction
solver in $rz$-geometry. It tests the time evolution of the temperature
profile using uniform and time independent heat conductivity.
In $rz$-geometry, the equation of the electron temperature for purely
heat conductive plasma follows
\begin{equation}
  C_{Ve}\frac{\partial T_e}{\partial t} = \frac{1}{r}
  \frac{\partial}{\partial r}\left( rC_e \frac{\partial T_e}{\partial r}
  \right) + \frac{\partial}{\partial z}\left(
  C_e \frac{\partial T_e}{\partial z} \right).
\end{equation}
We will set the electron specific heat $C_{Ve}=1$ and assume the electron
conductivity $C_e$ to be constant. In that case, a solution can be written as
a product of a Gaussian profile in the $z$-direction and an elevated Bessel
function $J_0$ in the $r$-direction \citep{arfken1985}:
\begin{equation}
  T_e = T_{\rm min} + T_0 \frac{1}{\sqrt{4\pi C_e t}}
  e^{-\frac{z^2}{4C_e t}} J_0(b r)e^{-b^2 C_e t},
\end{equation}
where $b\approx 3.8317$ is the first root of the derivative of
$J_0(r)$. We select the following values for the input parameters:
$T_{\rm min}=3$, $T_0=10$, and $C_e=0.1$ in dimensionless units.

The computational domain is $-3<z<3$ and $0<r<1$ discretized with
$3\times 3$ grid blocks of $30\times 30$ cells each. In the region $-1<z<1$ and
$1/3<r<2/3$, we apply one level of mesh refinement. We impose a symmetry
condition for the electron temperature on the axis. On all other boundaries the
electron temperature is fixed to the time dependent reference solution. We
time evolve this heat conduction problem with a preconditioned conjugate
gradient method from time $t=1$ to the final time at $t=1.5$. The
Crank-Nicolson approach is used to achieve second order accurate time
integration.

\begin{figure}
\plotone{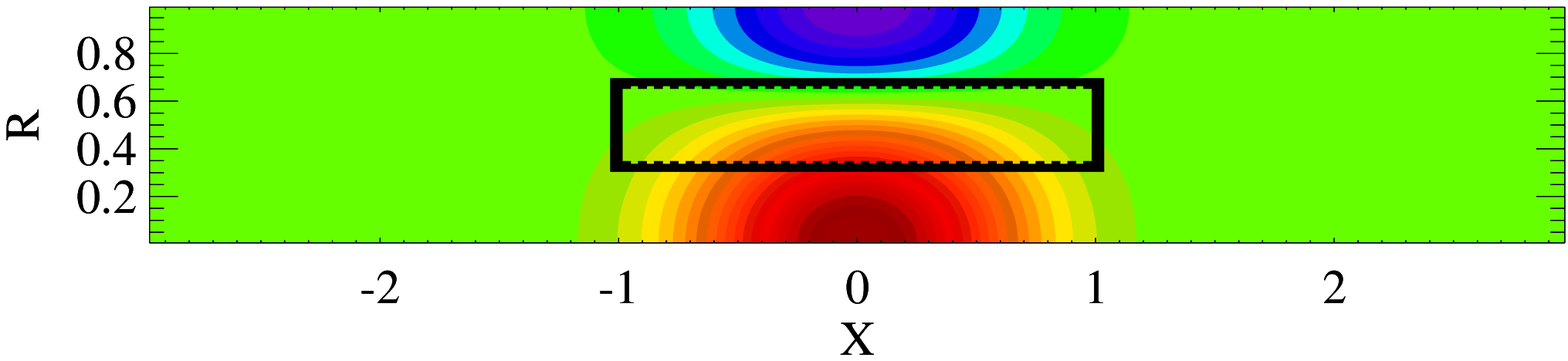}
\plotone{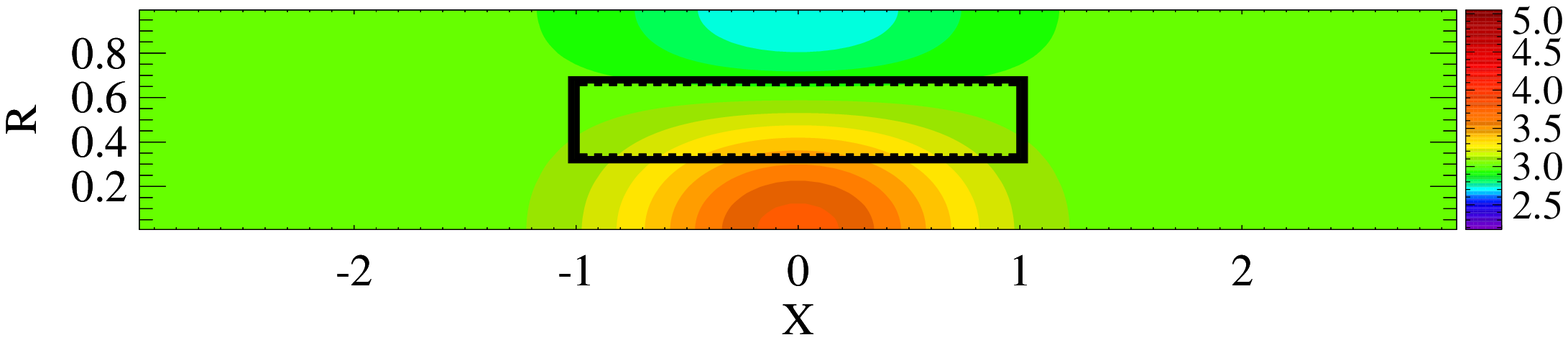}
\caption{The electron temperature for the uniform heat conduction test on a
non-uniform grid in $rz$-geometry. The top panel shows the electron temperature
in the initial condition while the bottom panel is the electron temperature at
the final time. The black box indicates the region within which the grid is
refined by one level.}
\label{fig:rz}
\end{figure}

\begin{figure}
\plotone{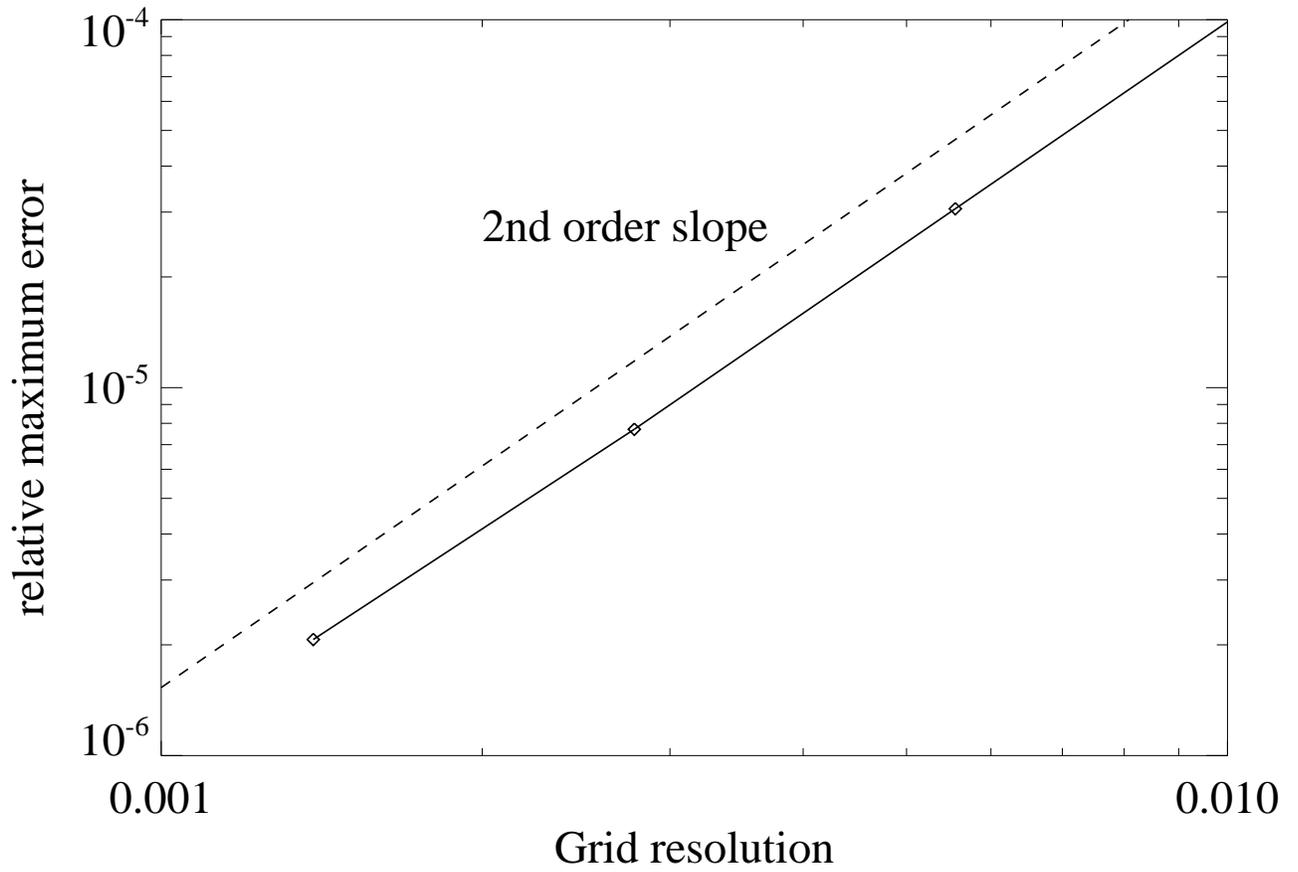}
\caption{The relative maximum error for the uniform heat conduction test on a
non-uniform grid in the $rz$-geometry.}
\label{fig:rzerror}
\end{figure}

The initial and final solutions for the electron temperature are shown in
Figure \ref{fig:rz} in color contour in the $rz$-plane. The heat conduction
has diffused the temperature in time to a more uniform state. The black
line indicates the region in which the mesh refinement was applied.
The relative maximum error of the numerically obtained electron temperature
versus the analytical solution is shown in Figure \ref{fig:rzerror}. Here we
used the non-uniform grid with base resolutions of $90^2$, $180^2$, $360^2$,
and $720^2$ cells and set the time step proportional to the cell size to
demonstrate a second order convergence rate.

\subsubsection{Reinicke Meyer-ter Vehn Test}

The \cite{rmtv1991} problem tests both the hydrodynamics and the heat
conduction implementation. This test generalizes the well-known
Sedov--Taylor strong point explosion in single temperature hydrodynamics by
including the heat conduction. The heat conductivity is parameterized as a
non-linear function of the density and material temperature:
$C_e = \rho^a T^b$. We select the spherically symmetric self-similar solution
of \cite{rmtv1991} with coefficients $a=-2$ and $b=13/2$ and the adiabatic
index is $\gamma=5/4$. This solution produces, similar to the Sedov--Taylor
blast-wave, an expanding shock front through an ambient medium. However, at
very high temperatures, thermal heat conduction dominates the fluid flows, so
that a thermal front precedes the shock front. With the selected parameters,
the heat front is always at twice the distance from the origin of the
explosion as is the shock front.

We perform the test in $rz$-geometry. The computational domain is divided
in $200\times 200$ cells. The boundary conditions along the $r$ and $z$ axes
are reflective. The two other boundaries, away from explosion, are prescribed
by the self-similar solution. The time evolution is numerically performed as
follows: For the hydrodynamics we use the
HLLE scheme and the generalized Koren limiter with $\beta=2$ as the
slope limiter. The CFL number is set to 0.8. The heat conduction is solved
implicitly with the preconditioned conjugate gradient method. The test
is initialized with the spherical self-similar solution with the shock front
located at the spherical radius $0.225$ and the heat front is at $0.45$. The
simulation is stopped once the shock front has reached $0.45$ and the heat
front is at $0.9$.

\begin{figure}
{\resizebox{\textwidth}{!}{\includegraphics[clip=]{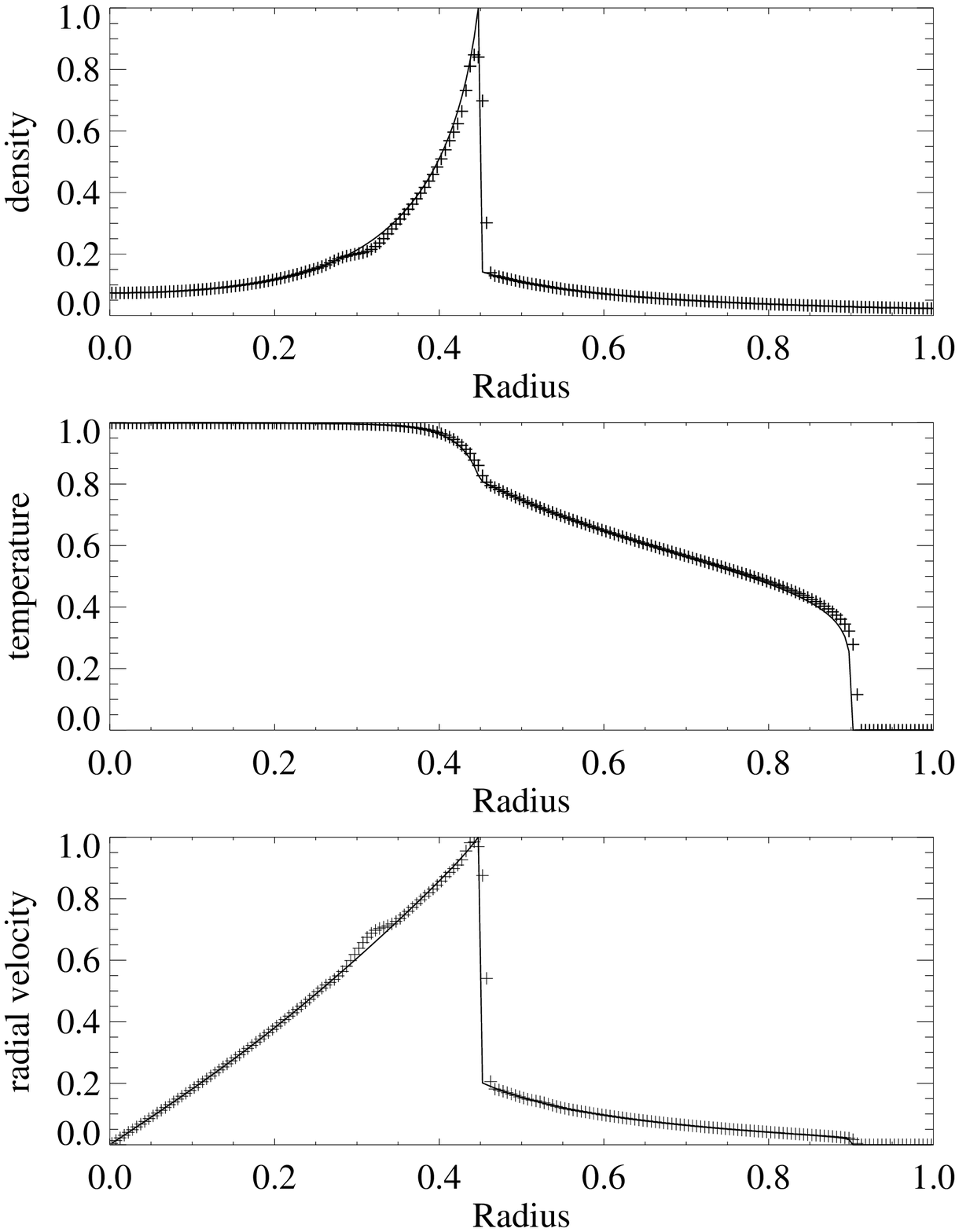}}} \\
{\resizebox{\textwidth}{!}{\includegraphics[clip=]{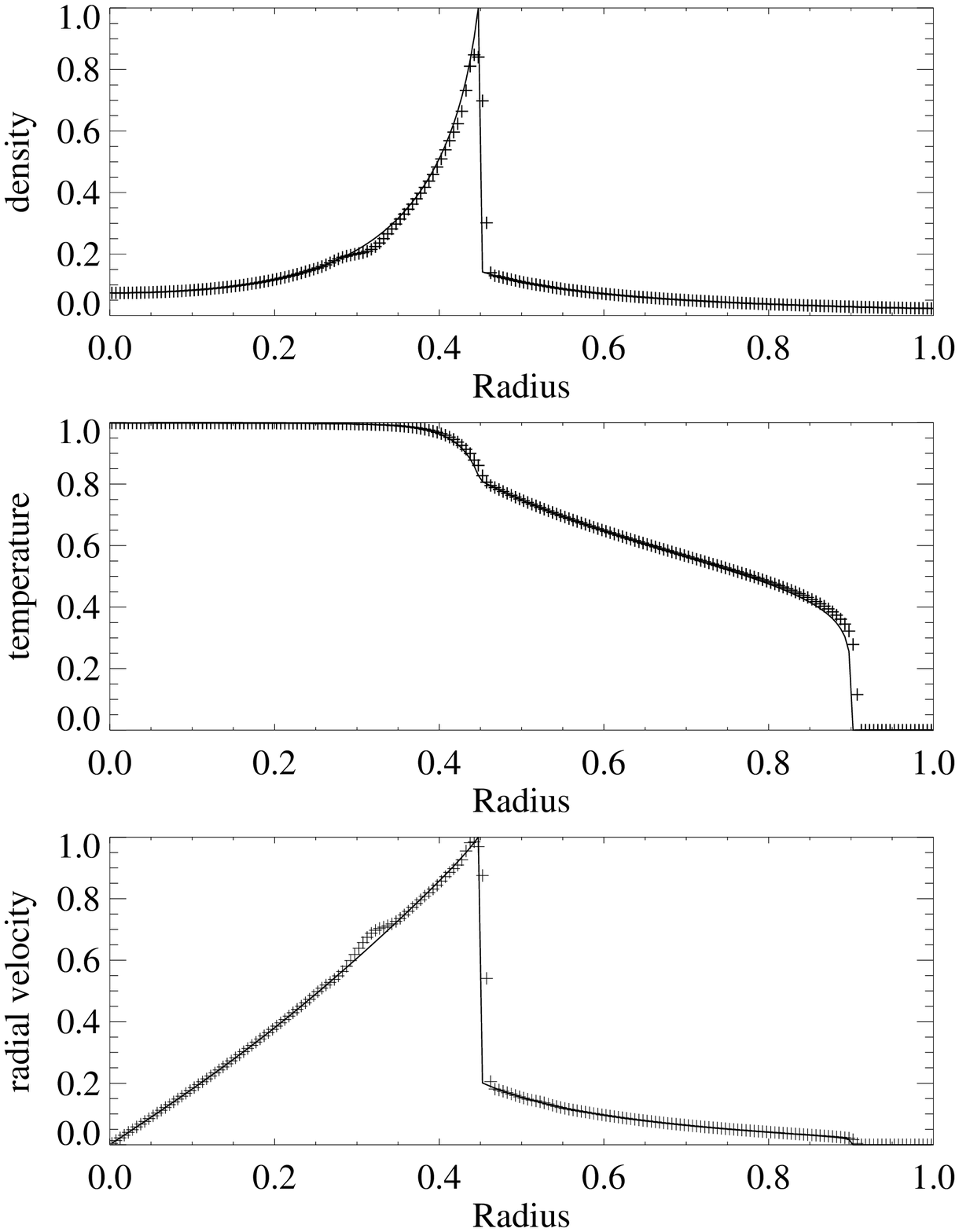}}} \\
{\resizebox{\textwidth}{!}{\includegraphics[clip=]{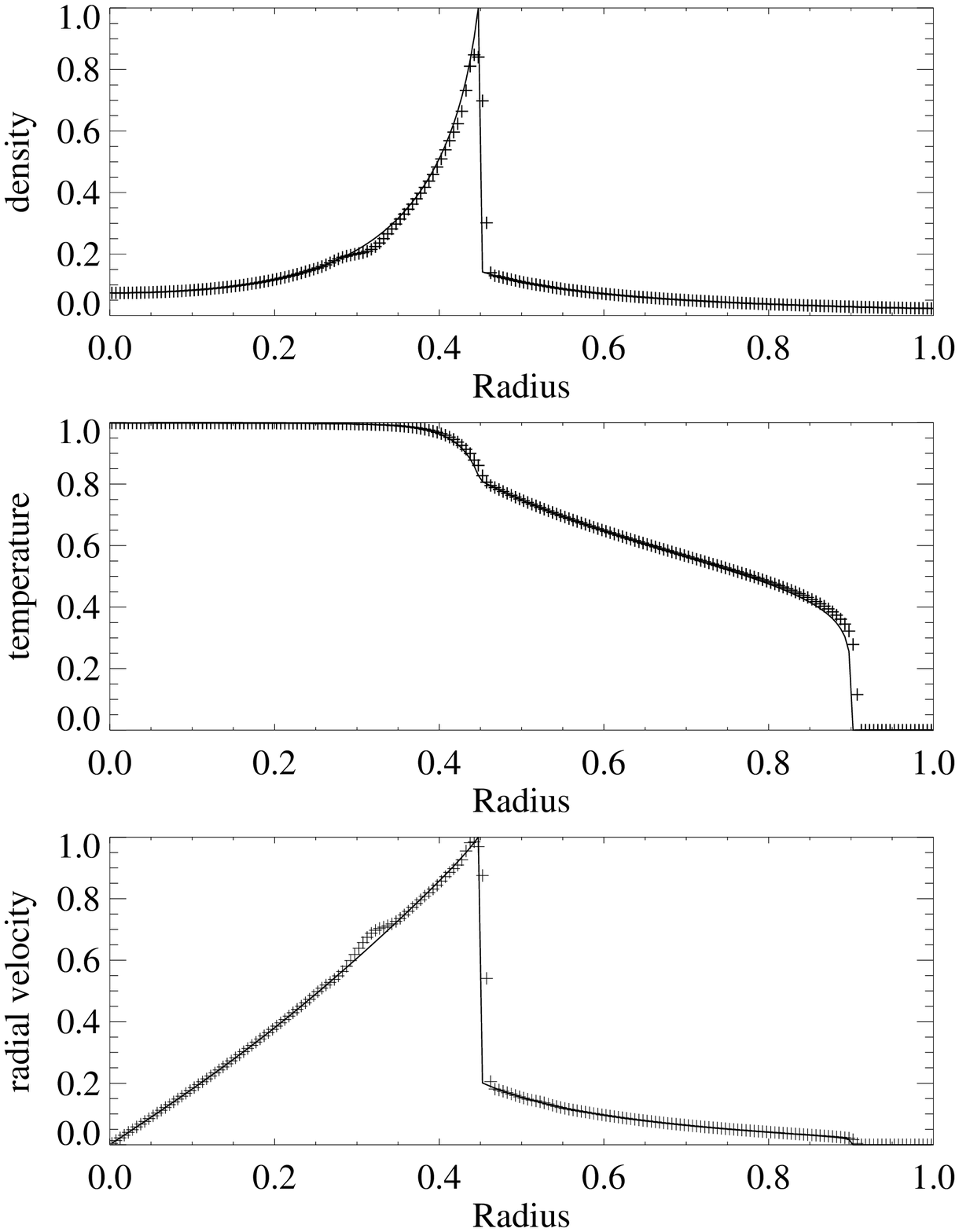}}}
\caption{Density (top panel), temperature (middle panel), and radial velocity
(bottom panel) along the $z=0$ cut for the Reinicke Meyer-ter Vehn test in
$rz$-geometry. The numerical solution (+ symbols) is at the final time
compared to the self-similar analytical reference solution (solid lines).}
\label{fig:rmtvcut}
\end{figure}

\begin{figure}
\plotone{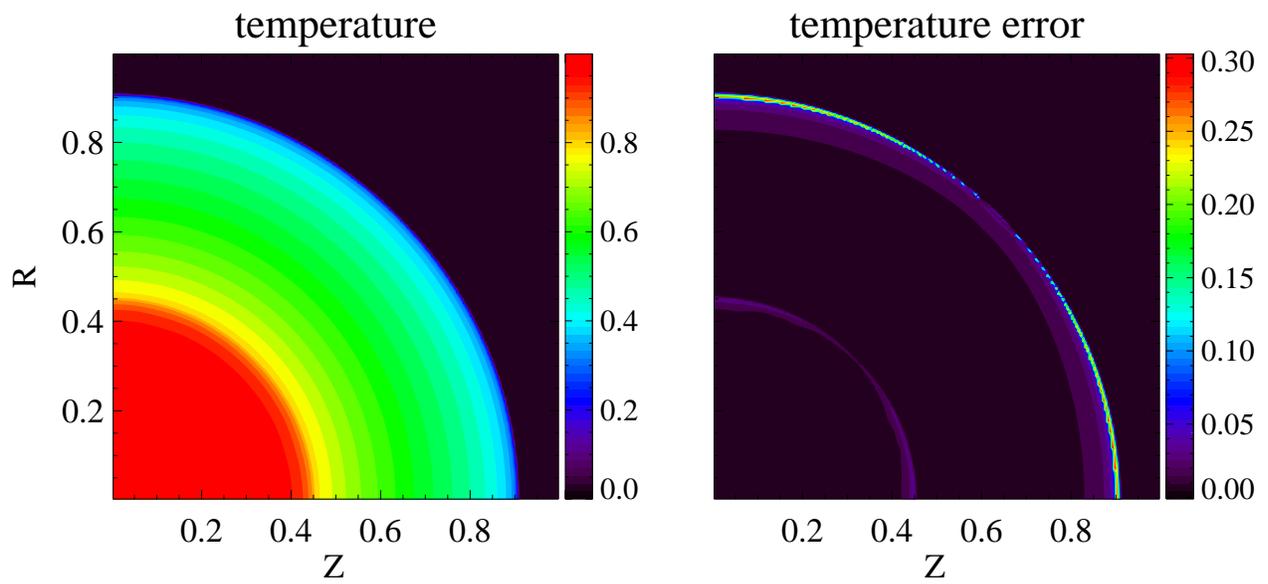}
\caption{The temperature (left panel) and temperature error compared to the
reference solution (right panel) for the Reinicke Meyer-ter Vehn test in
$rz$-geometry.}
\label{fig:rmtvtemp}
\end{figure}

A 1D slice along the $r$-axis of the solution at the final time is shown in
Figure \ref{fig:rmtvcut}. We normalize the output similar to \cite{rmtv1991}:
The temperature is normalized by the central temperature, while the density
and radial velocity are normalized by their values of the post-shock state
at the shock. The numerical solution obtained by the CRASH code is shown as
$+$ symbols and is close to the self-similar reference solution, shown as
solid lines. Note that the temperature is smooth due to the heat conduction,
except for the discontinuous derivative at the heat front. The wiggle at
$r=0.3$ in the density and radial velocity is a due to the diffusion of the
analytical shock discontinuity in the initial condition during the first few
time steps. In the left panel of Figure \ref{fig:rmtvtemp}, the spherical
expansion of temperature at the final time is shown. Clearly, the Cartesian
grid with the $rz$-geometry does not significantly distort the spherical
symmetry of the solution. The spatial distribution of the error in the
temperature is shown in the right panel. The errors are largest at the
discontinuities of the shock and heat fronts as expected.

\begin{figure}
\plotone{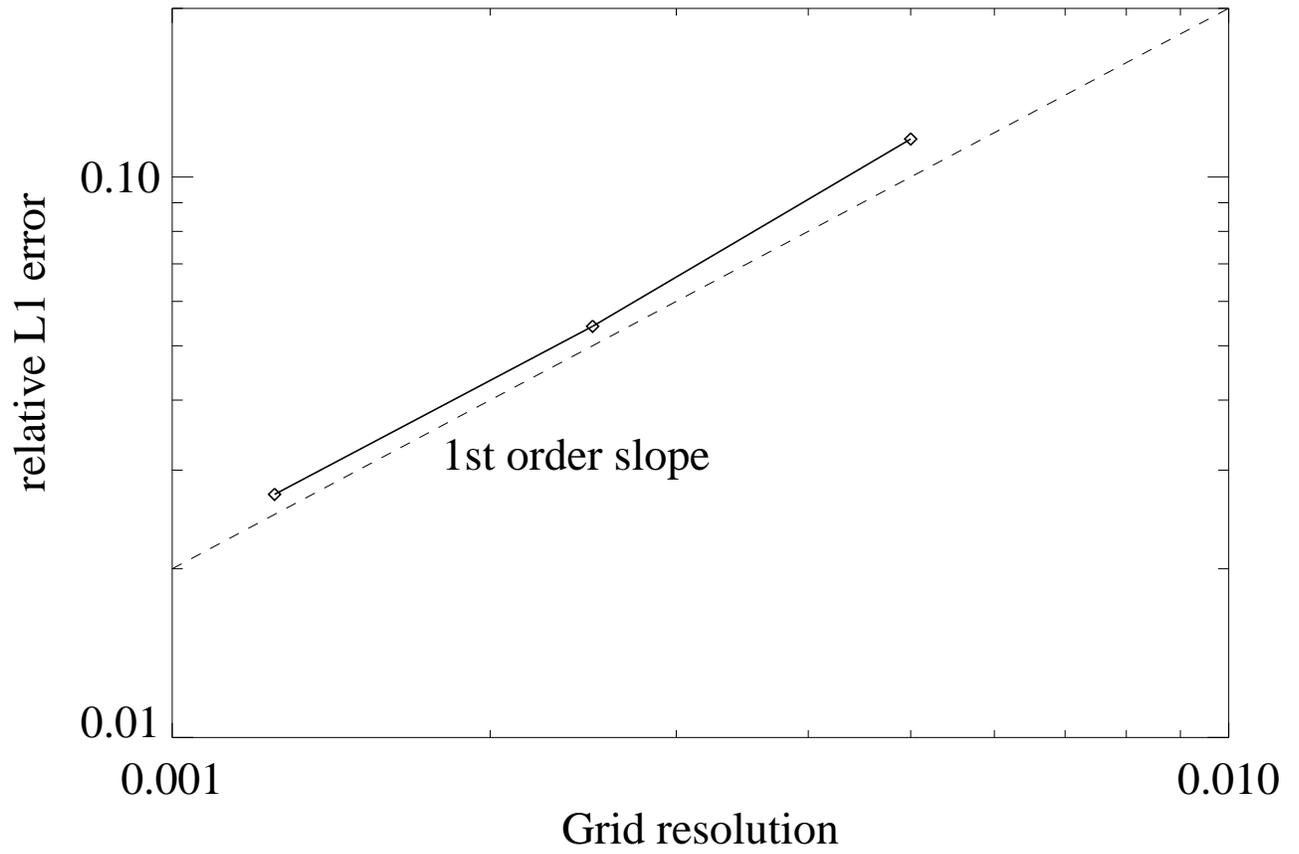}
\caption{The relative L1 error for the Reinicke Meyer-ter Vehn test in
$rz$-geometry.}
\label{fig:rmtverror}
\end{figure}

A grid convergence study is performed with resolutions of $200^2$, $400^2$, and
$800^2$ cells. The relative L1 error in Figure \ref{fig:rmtverror} is
calculated using the density, velocity components, and the material
temperature. The convergence rate is first order due to the shock and heat
front.

\subsubsection{Heat Conduction Version of Lowrie's Test}
\label{sec:lowrie3}

Any of the verification tests for non-equilibrium gray-diffusion coupled to the
single temperature hydrodynamics can be reworked as a test for the
hydrodynamic equations for the ions coupled to the electron pressure equation
with electron heat conduction and energy exchange between the electrons and
ions. As an example, we will transform one of the non-equilibrium gray
diffusion tests of \cite{lowrie2008} to verify the heat conduction
implementation.

The electron energy density equation (\ref{eq:electronenergy}) without the
radiation interaction can be written as
\begin{equation}
  \frac{\partial E_e}{\partial t} + \nabla\cdot\left[ E_e \bu \right]
  + p_e\divu = \nabla\cdot\left[ C_e\nabla T_e \right]
  + \sigma_{ie} (T_i - T_e),\label{eq:electrontemp}
\end{equation}
where the heat conduction and energy exchange terms on the right hand side
depend on the gradients and differences of the temperatures. The equation for
the gray radiation energy density (\ref{eq:grayenergy}) on the other hand
depends on the gradients and differences of energy densities. By defining
the radiation temperature $T_r$ by $E_r=aT_r^4$ and using the definition of the
Planckian $B=aT_e^4$, we can rewrite the energy density equation for the
radiation as
\begin{eqnarray}
  \frac{\partial E_r}{\partial t} + \nabla\cdot \left[ E_r\bu \right]
  + \frac{1}{3} E_r \divu &=& \nabla\cdot\left[ D_r \nabla E_r \right]
  + c\kappa_P (aT^4 - E_r) \nonumber \\
  &=& \nabla\cdot\left[ \overline D_r \nabla T_r \right]
  + c\overline\kappa_P (T-T_r),\label{eq:radiationtemp}
\end{eqnarray}
where $\overline D_r = D_r 4aT_r^3$ and
$c\overline\kappa_P = c\kappa_Pa(T^2+T_r^2)(T+T_r)$ are the new coefficients
that appear due to this transformation.
The equations (\ref{eq:electrontemp}) and (\ref{eq:radiationtemp}) are now
of the same form. To translate a gray diffusion test to a heat conduction test,
we reinterpret $\overline D_r$ as the heat conductivity $C_e$ and
$c\overline\kappa_P$ as the relaxation coefficient $\sigma_{ie}$ in the
ion-electron energy exchange. In addition, the material temperature $T$ and
radiation temperature $T_r$ have to be reinterpreted as the ion temperature
$T_i$ and electron temperature $T_e$, respectively. Note that we also have to
relate the electron pressure and internal energy by $p_e=E_e/3$ similar to
the radiation field corresponding to $\gamma_e=4/3$, and let the electron
internal energy and electron specific heat depend on the electron temperature
as $E_e=aT_e^4$ and $C_{Ve}=4aT_e^3$, respectively.

As an example, we transform the Mach 5 non-equilibrium gray diffusion
shock tube problem of \cite{lowrie2008}.
It uses non-uniform opacities that depend on the density and temperature
defined by $D_r=0.0175(\gamma T)^{7/2}/\rho$ and $c\kappa_P = 10^6/D_r$. The
above described procedure is used to translate this problem to an electron
heat conduction test with energy exchange between the electron and ions.
The heat conductivity for this test is
$C_e=4aT_e^3 0.0175(\gamma T_i)^{7/2}/\rho$ and the relaxation coefficient
between the electron and ions is
$\sigma_{ie}=a(T_i^2+T_e^2)(T_i+T_e)4aT_e^3 10^6/C_e$.

\begin{figure}
\begin{center}
{\resizebox{0.48\textwidth}{!}{\includegraphics[clip=]{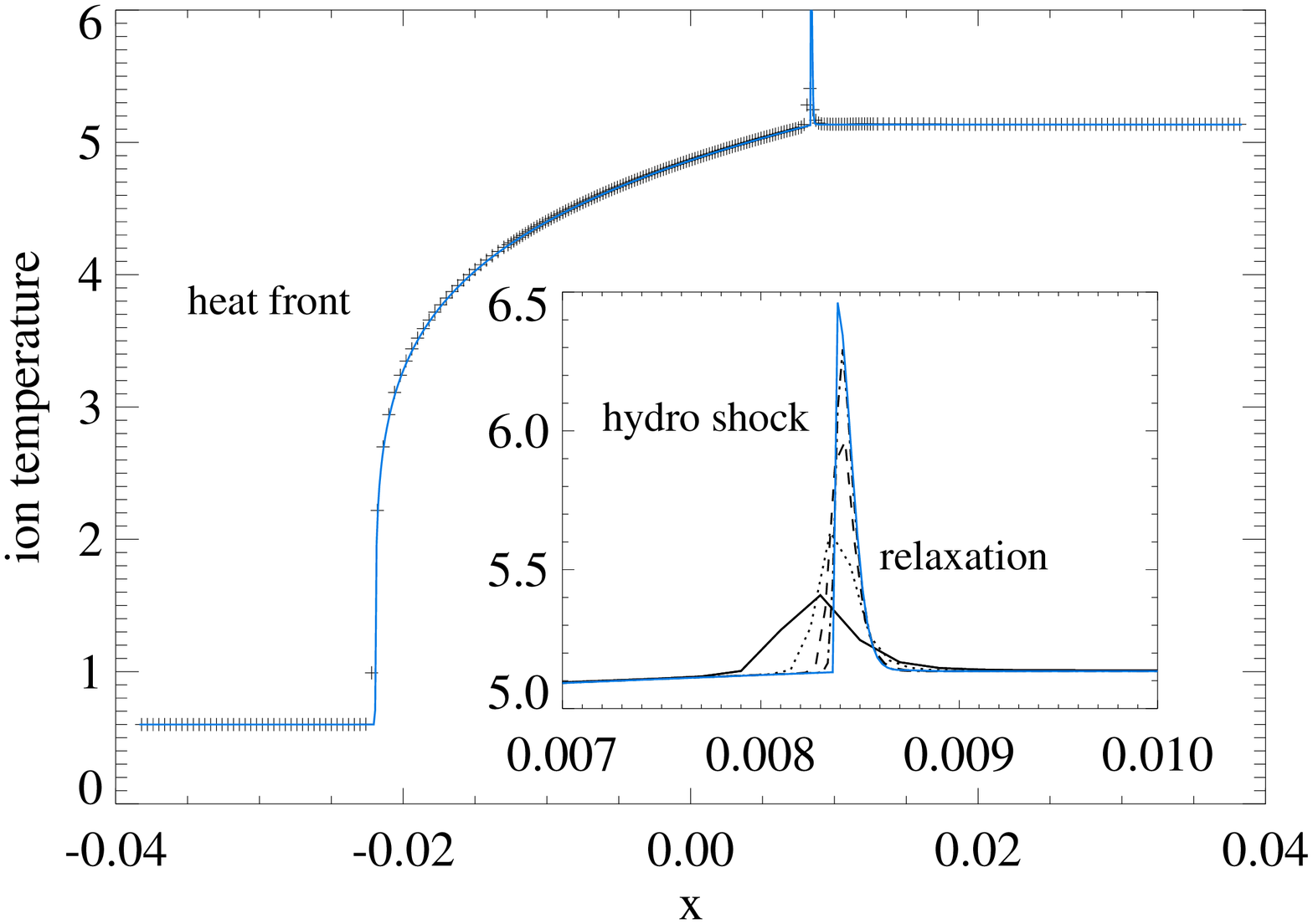}}}
{\resizebox{0.48\textwidth}{!}{\includegraphics[clip=]{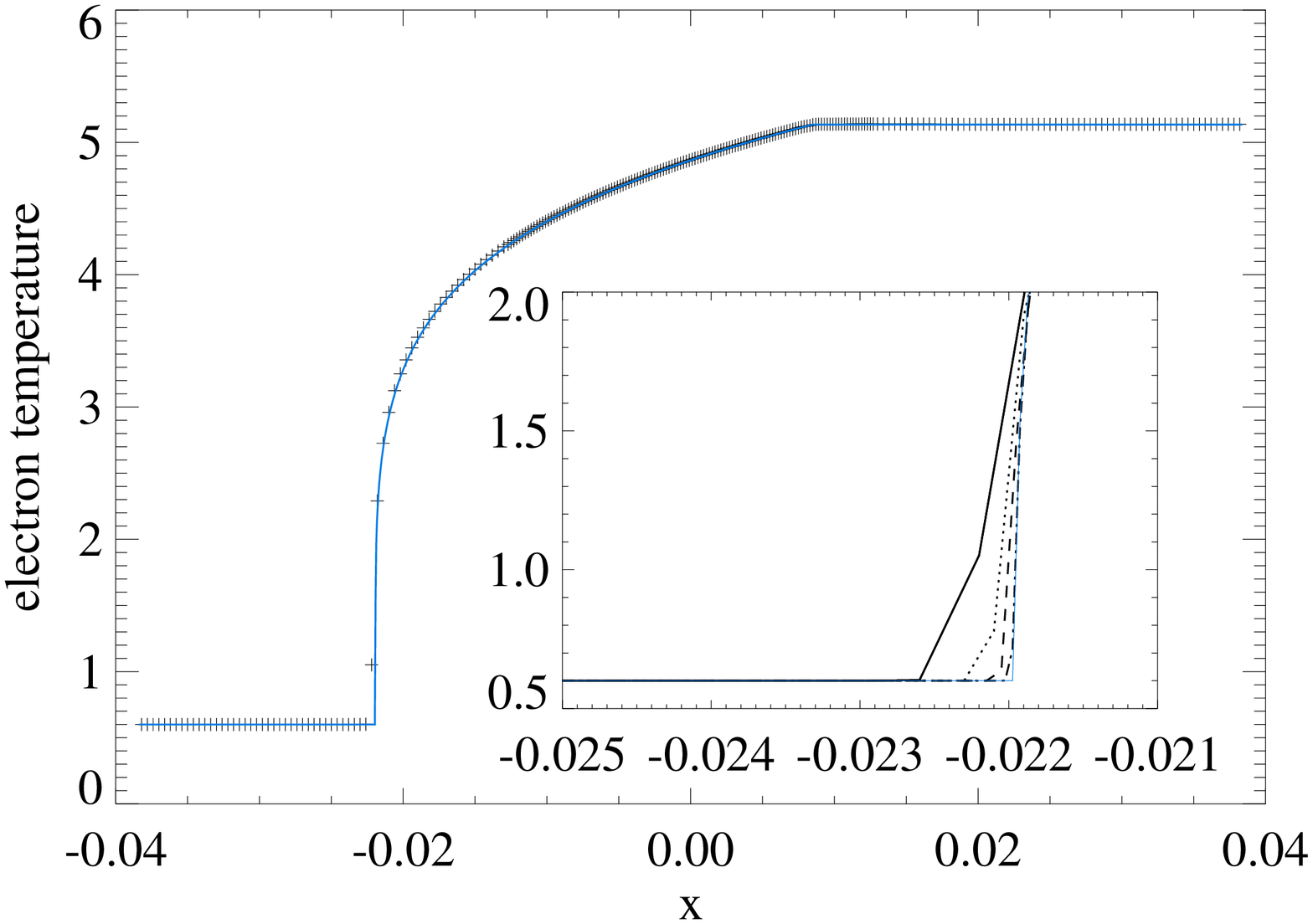}}}
\end{center}
\caption{Mach 5 shock tube problem of \citet{lowrie2008} transformed
to a non-uniform heat conduction and ion-electron collison frequency test
and rotated on a 2D non-uniform grid. The ion (left panel)
and electron (right panel) temperatures at the final time are shown in the
$x$-direction. The blue line is the reference solution. In the left
panel, the grid convergence near the shock is shown in the inset. In the
right panel, a blow-up of the grid convergence to the reference heat front
is shown.}
\label{fig:lowrie3temp}
\end{figure}

We perform this Mach 5 heat conduction test on a 2D non-uniform grid. For
the initial condition, the 1D semi-analytical steady state reference solution
of \cite{lowrie2008} is used. There is
a Mach 5 pre-shock flow on the left side of the tube resulting in an embedded
hydro shock as well as a steep thermal front (a look at Figure
\ref{fig:lowrie3temp} will help to understand this shock tube problem.)
We add an additional velocity of
Mach $-5$ so that the pre-shock velocity is zero and the shock is no longer
steady, but instead will move to the left with a velocity $-5$ (in units
in which the pre-shock speed of sound is 1). The problem
is rotated anti-clockwise on the grid by $\tan^{-1}(1/2)$. The
translational symmetry is now in the $(-1,2)$ direction in the $xy$-plane
similar to the Mach 1.05 shock tube problem described in Section
\ref{sec:lowrie1}. The computational domain is $-0.0384<x<0.0384$ by
$-0.0048<y<0.0048$. Inside the area $-0.0128<x<0.0128$ and $-0.0016<y<0.0016$,
we apply one level of refinement. This refinement is set up such that
both the heat front as well as the shock front will
propagate through the resolution change on the left (from fine to coarse)
and right (from coarse to fine), respectively. For the boundary conditions in
the $x$-direction, we fix the state on the
right side with the semi-analytical solution, but for the left side we
use zero gradient. On the $y$ boundaries, we
apply a sheared zero gradient in the $(-1,2)$ direction.

For the evolution till the final time $t=0.0025$, we use the HLLE scheme
together with the generalized Koren limiter with $\beta=3/2$ to solve the
hydrodynamic equations. The CFL number is set to 0.8. The heat conduction and
energy exchange between electrons and ions are solved implicitly with the
backward Euler scheme using the GMRES iterative solver in combination with a
BILU preconditioner.

In Figure \ref{fig:lowrie3temp}, the electron (right panel) and ion (left
panel) temperatures are shown at the final time along the $x$-axis. The
semi-analytical reference solution is shown as a blue line, while the
numerical solution is shown with $+$ symbols for a simulation with
$192\times 24$ cells at the base level in the $x$ and $y$ direction.
The hydro shock is located near $x\approx 0.0085$ and shows up in the ion
temperature as a jump in the temperature, followed directly behind the shock by
a strong relaxation due to the energy exchange between the ions and electrons.
The electron temperature stays smooth due to strong heat conduction.
The heat front is seen with a steep foot at $x\approx -0.022$. This front
corresponds to the radiative precursor in the non-equilibirum gray diffusion
tests of \cite{lowrie2008}. We repeated the test with four different
resolutions at the base level: $192\times 24$, $384\times 48$, $768\times 96$,
and $1536\times 192$ cells in the $x$ and $y$ direction. The insets in both
panels of Figure \ref{fig:lowrie3temp} show the four
resolutions as solid, dotted, dashed, dashed-dotted lines, respectively.
In the left panel, the zoom-in shows the convergence of the ion temperature
towards the embedded hydro shock and the temperature relaxation. In the right
panel, the blow-up shows the convergence towards the reference precursor
front. Note that no spurious oscillations appear near the shock or near the
precursor.

\begin{figure}
\plotone{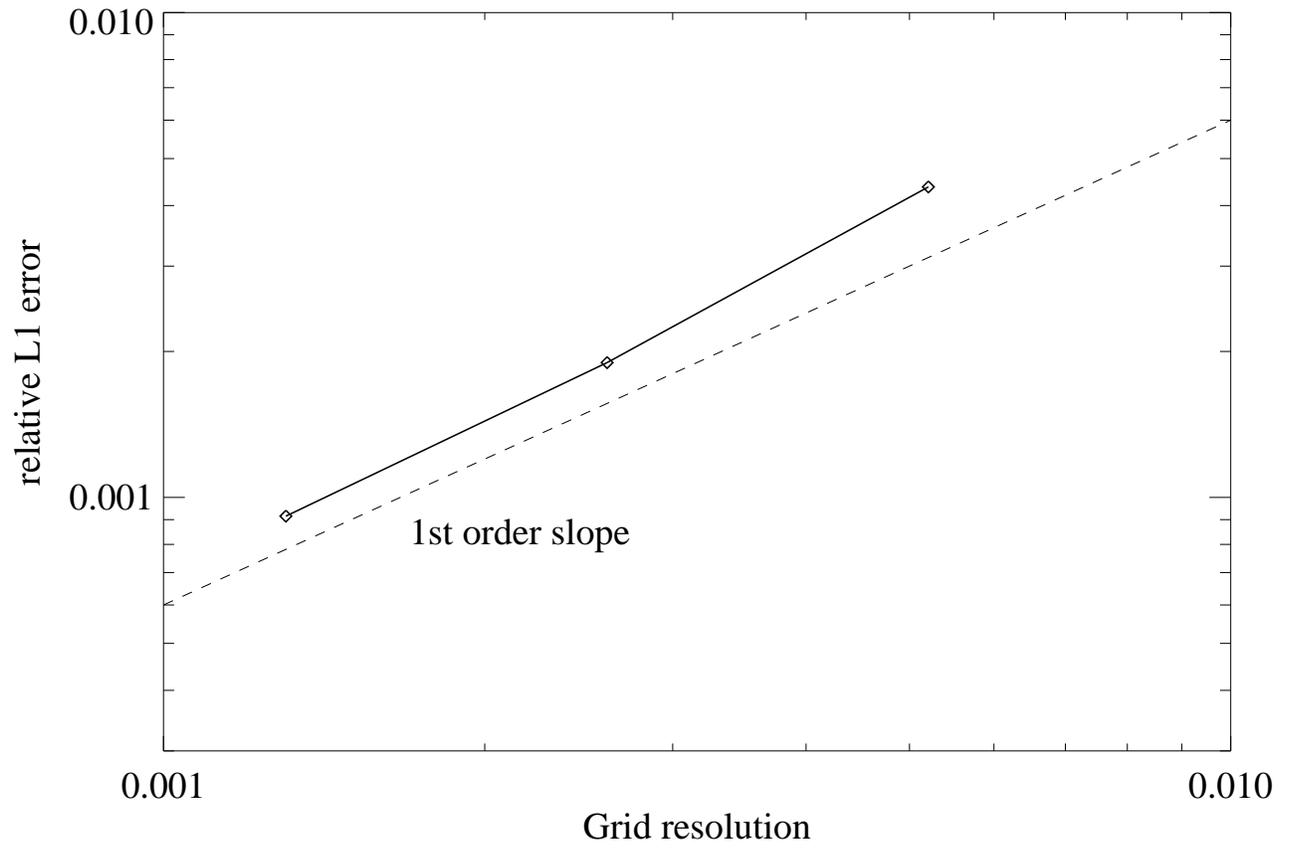}
\caption{Relative L1 error for the Mach 5 non-equilibrium heat conduction
test on a non-uniform grid.}
\label{fig:lowrie3error}
\end{figure}

Due to the discontinuity in both the shock and heat precursor, the convergence
rate can be at most first order. Indeed, in Figure \ref{fig:lowrie3error}
the relative L1 error shows first order accuracy. The error
is calculated using all the density, velocity components and both
temperatures. Note that the spike in the ion temperature is spatially so
small that a huge number of grid cells are needed to get a fully resolved
shock and relaxation state.

\subsection{Full System Tests}\label{sec:fullsystem}

\begin{figure}
\begin{center}
{\resizebox{0.48\textwidth}{!}{\includegraphics[clip=]{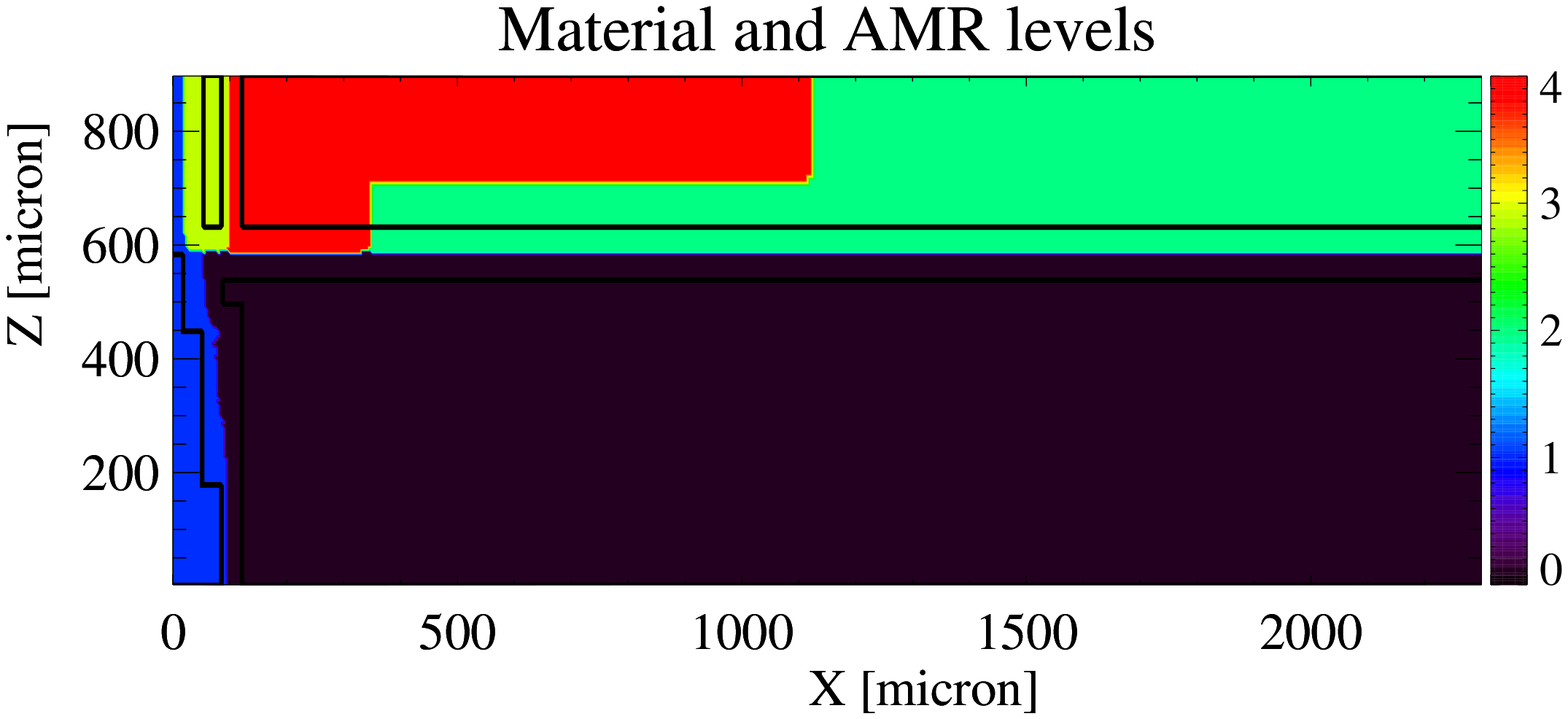}}}
{\resizebox{0.48\textwidth}{!}{\includegraphics[clip=]{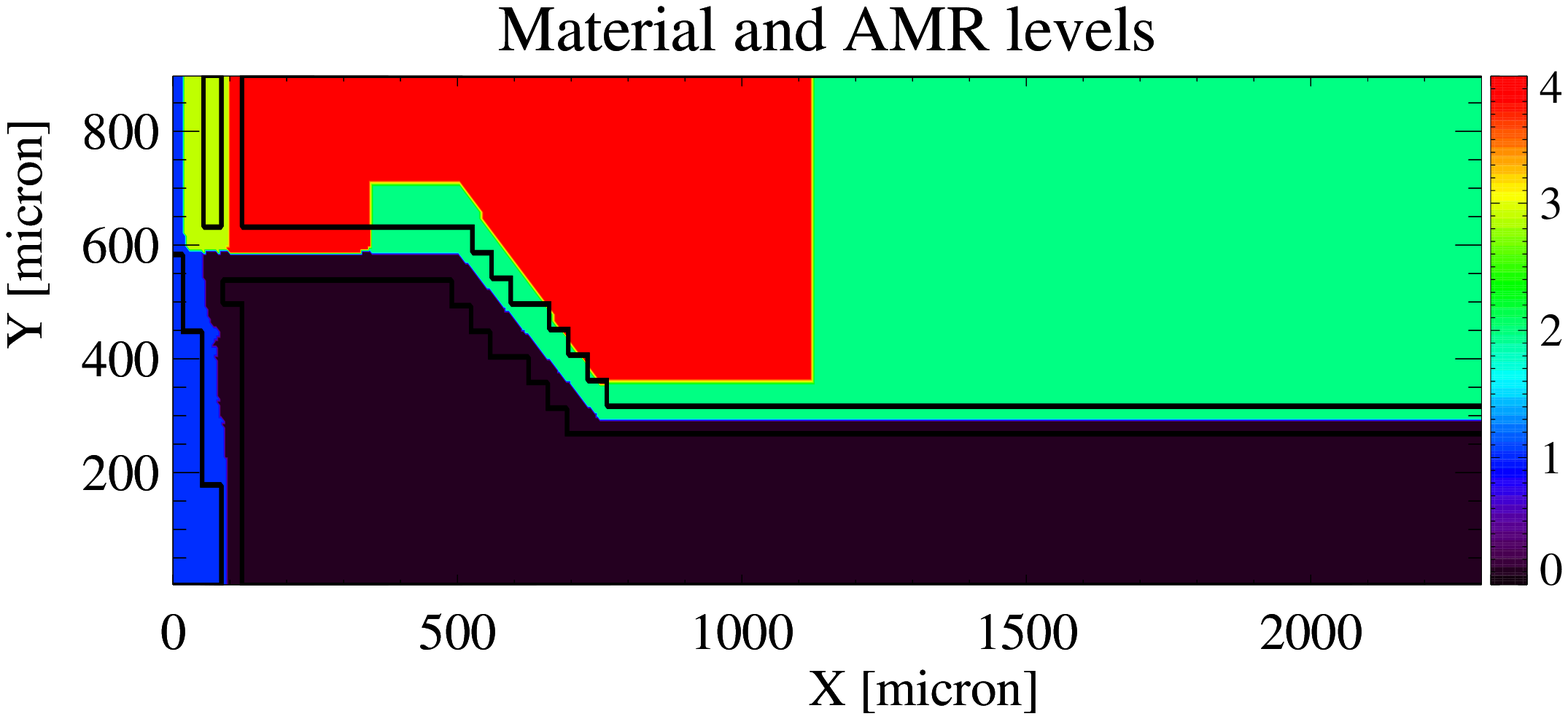}}}
\end{center}
\caption{The geometry of the 3D elliptic nozzle experiment after 1.1\,ns,
consisting of 5 materials: Beryllium (blue), Xenon (black), polyimide (green),
gold (yellow), and acrylic (red) in both panels. The radius of the inside of
the polyimide tube is 600\,$\mu$m in the $y=0$ plane (left panel). In the
$z=0$ plane (right panel), the radius of the inner tube is 600\,$\mu$m for
$x<500$\,$\mu$m, but shrinks to 300\,$\mu$m beyond $x=750$\,$\mu$m. The lines
represent the mesh refinement at material interfaces and shock fronts.}
\label{fig:nozzle_init}
\end{figure}

The CRASH test repository contains a range of full system configurations
to be used for validation with future laboratory experiments. In Figure
\ref{fig:nozzle_init}, we show the configuration of a 3D elliptic nozzle
through which a fast shock of the order of $150$\,km/s will be launched, which
is still significantly slower than the speed of light. The shock wave is
produced by a 1.1\,ns laser pulse from the
left with 4\,kJ of energy irradiating a 20\,$\mu$m thick Beryllium
disk, initially located at $x=0$. A layer of gold is glued to
the plastic tube to protect the outside of the tube from the laser-driven
shock. The plastic (polyimide) tube is circular for
$x<500$\,$\mu$m with a radius of 600\,$\mu$m. Beyond $x=750$\,$\mu$m the
tube is elliptic by flattening the tube in the $y$-direction by a factor 2.

The first part of the simulation is performed with the 2D, Lagrangian,
radiation hydrodynamic code HYADES \citep{larsen1994} to time advance the
laser energy deposition and the response of the system until the end of the
laser pulse at 1.1 ns. This laser pulse first shocks and then accelerates the
Beryllium to the right. After 1.1\,ns the output of HYADES is used as an
initial condition of the CRASH code.

This simulation is performed for a two temperature, electron and ion, plasma.
For the radiation, we use the flux limited diffusion approximation with 30
groups. The photon energy is in the range of 0.1\,eV to 20\,keV,
logarithmically distributed over the groups. Due to the symmetry in the
problem we only simulate one quadrant ($y>0$ and $z>0$), with reflective
boundary conditions at $y=0$ and $z=0$. At all other boundaries we use an
extrapolation with zero gradient for the plasma and a zero incoming flux
boundary for the radiation. The domain size is
$\left[ -150, 3900 \right] \times \left[ 0,900 \right] \times
\left[ 0,900 \right]$ microns for the $x$, $y$, $z$ coordinates.
The base level grid consists of $120\times20\times20$ blocks of
$4\times4\times4$ mesh cells. One level of dynamic mesh refinement is used at
material interfaces and the shock front.
Overall, the effective resolution is $960\times160\times160$ cells and
there are approximately 4.5 million finite volume cells.
The hydrodynamic equations are solved with the HLLE scheme with a CFL
number 0.8 together with the generalized Koren limiter with $\beta=3/2$.
The diffusion and energy exchange of the radiation groups as well as
the heat conduction are solved with the decoupled implicit scheme
using a Bi-CGSTAB iterative solver. The simulation from 1.1\,ns to 13\,ns
physical time took 1 hour and 55 minutes on 480 cores of the FLUX
supercomputer at the University of Michigan.

\begin{figure}
\plotone{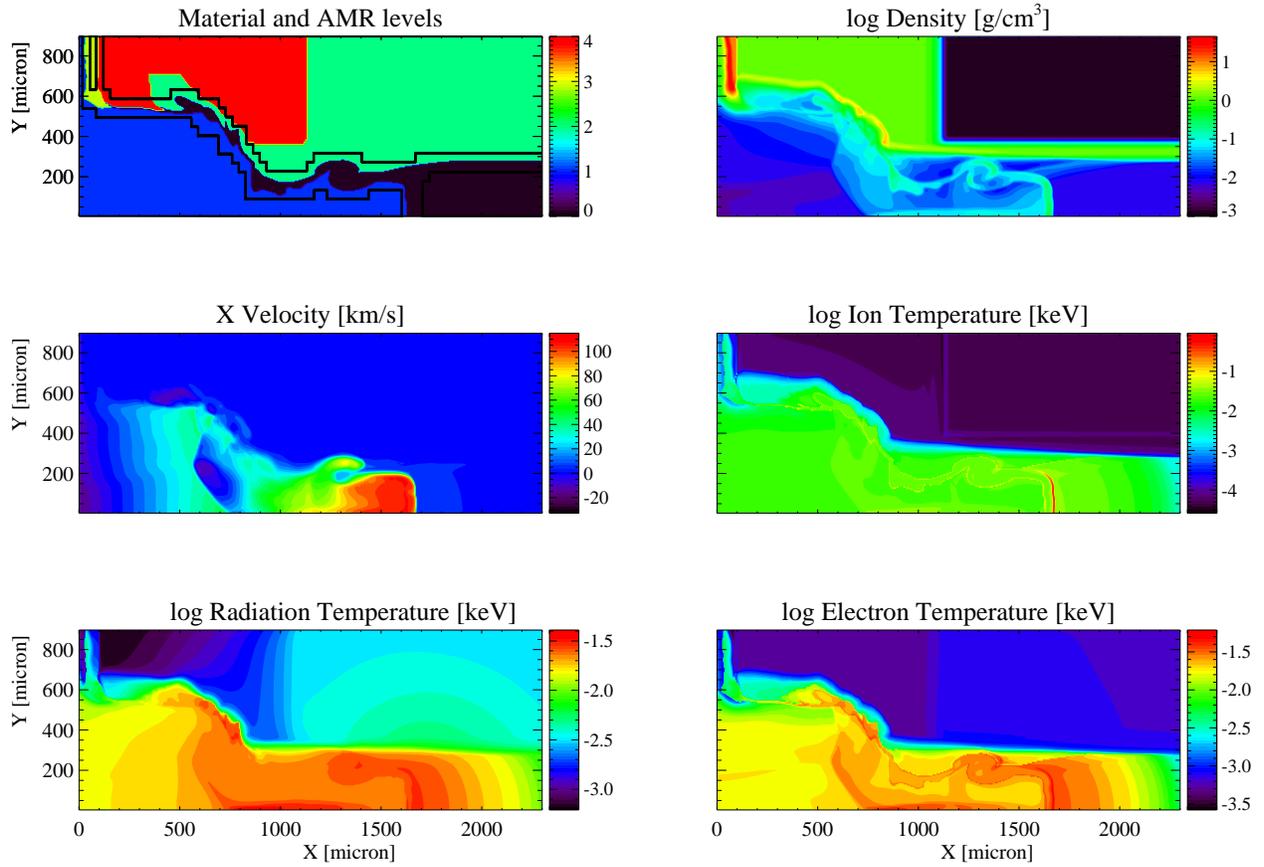}
\caption{Simulated radiative shock structure at 13ns in a 3D elliptic nozzle
consisting of the 5 materials indicated in Figure \ref{fig:nozzle_init}.
The plots show in the $xy$-plane in color contour the variables indicated in
the plot title. The primary shock is at $x\approx 1700$.}
\label{fig:nozzle}
\end{figure}

In Figure \ref{fig:nozzle}, we show the shock structure at 13\,ns. The
accelerated Beryllium compresses the Xenon directly to the right of the
interface, which
is seen as a high density plasma near $x=1700$\,$\mu$m in the top right panel
of Figure \ref{fig:nozzle}. This drives a primary shock and the velocity jump
at $x\approx1700$\,$\mu$m is seen in the middle left panel. Behind the shock
front, the ions are heated as depicted from the middle right panel, followed
directly behind the shock by a cooling due to the energy exchange between the
ions and electrons. Early on, the electron heating produces ionization and
the emission of radiation, and the radiation in turn heat and ionizes the
material ahead of the primary shock. 
The radiation temperature, measuring the total radiation energy
density, is shown in the bottom left panel. The photons will interact again
with the matter, sometimes after traveling some distance. This is the source
of the wall shock seen ahead of the primary shock \citep{doss2009,doss2011}:
photons traveling ahead of the shock interact with the
plastic wall, heat it, and this in turn drives a shock off the wall into the
Xenon. The ablation of the plastic is depicted in the top left panel as a
radially inward moving polyimide (in green color) near and even ahead of the
primary shock. The compressed Xenon due to the plastic ablation is seen in
the top right panel as a faint density feature that is ahead of the primary
shock front, between $x=1700$\,$\mu$m and $x=2000$\,$\mu$m.
The interaction between the photons and matter is also seen by
the radiative precursor to the right of the radiative shock elevating
the electron temperature ahead of the shock in the
bottom right panel. This is due to the strong coupling between the electrons
and radiation field. The reader is referred to \citet{drake2011} for more
details on radiative effects in radiative shock tubes.

\subsection{Parallel Performance}\label{sec:parallel}

\begin{figure}
\plotone{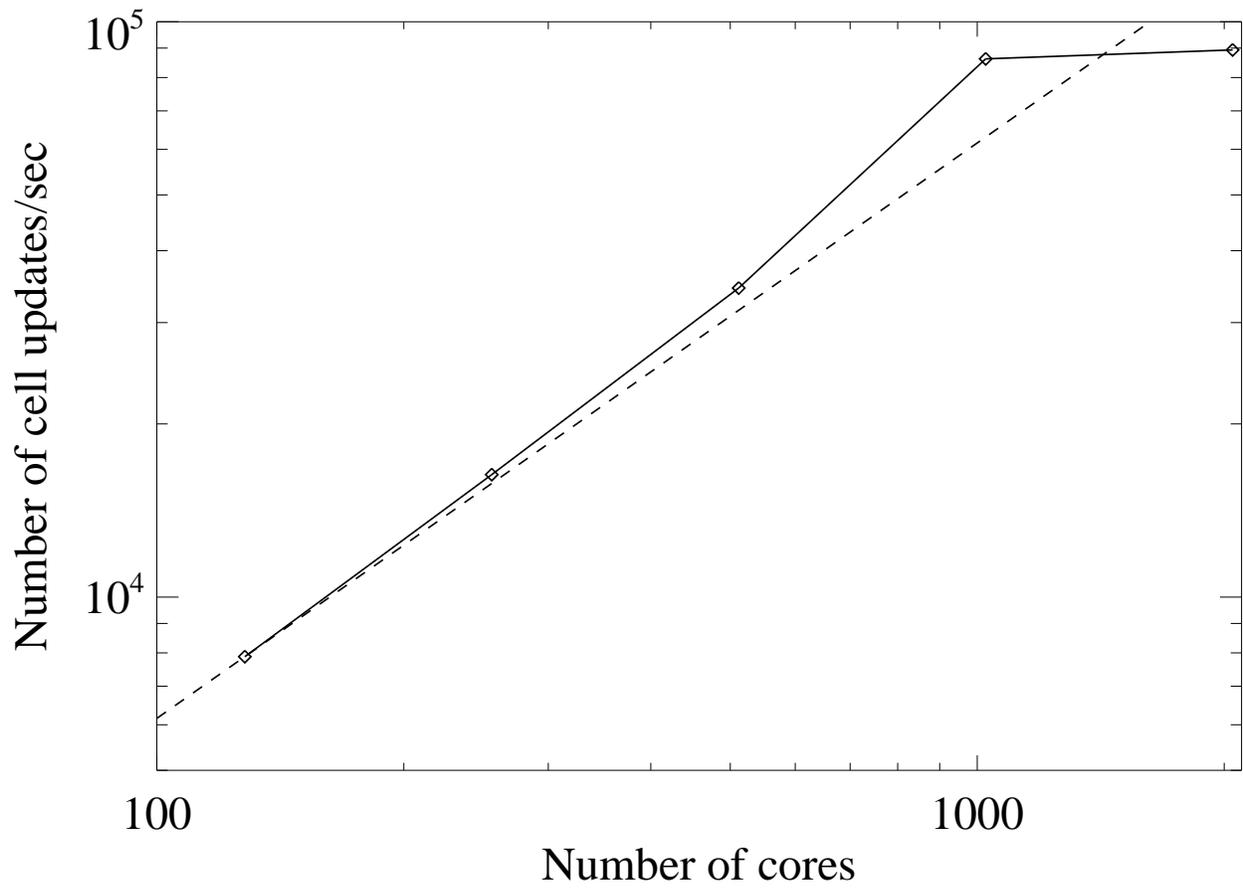}
\caption{Strong scaling of the CRASH code, running a 3D CRASH application with
5 material level sets, electron and ion temperature, 30 radiation groups, and
two levels of time dependent mesh refinement.}
\label{fig:scaling}
\end{figure}

We present parallel scaling studies on the Pleiades supercomputer at NASA
Ames. This computer is an SGI ICE cluster connected with infiniband.
Figure \ref{fig:scaling} shows the strong scaling for a problem size that
is independent of the number of processors. This 3D
simulation is a circular tube version of the full system test described in
Section \ref{sec:fullsystem}. It uses five materials, 30 radiation groups,
and separate electron and ion temperatures. The grid contains
$80\times8\times8$ blocks of $4\times4\times4$ cells each at the
base level and in addition two time dependent refinement
levels. There are overall approximately 2.6 million cells in this problem.
We use lookup tables for the EOS and opacities, so that the computational
time for that is negligible. For the hydrodynamic equations, we use the HLLE
scheme together with the generalized Koren limiter with $\beta=3/2$. The
radiation diffusion, electron heat conduction and energy exchange terms are
solved implicitly with the decoupled scheme, uing the Bi-CGSTAB iterative
solver. This simulation is performed for 20 time steps for the number of
cores varying from 128 to 2048, but excludes file I/O to measure the
performance of the implicit solver. Up to 1024 cores, we get good scaling.
However for more cores we observe saturation in the performance.

\section{SUMMARY}\label{sec:summary}

We have extended the BATS-R-US code \citep{powell1999,toth2010} with a new
radiation transfer and heat conduction library. This new combination together
with the equation-of-state and multigroup opacity solver is called the
CRASH code. This code uses the recently developed parallel Block Adaptive
Tree Library (BATL, see \citet{toth2010}) to enable highly resolved
radiation hydrodynamic solutions. The implemented radiation hydrodynamic
schemes solve for the gray or multigroup radiation diffusion models in the
flux limited diffusion approximation.

In high energy density plasmas, the electrons are most of the time strongly
coupled to the ions by collisions. An important exception is at hydrodynamic
shocks, where the ions are heated by the shock wave and the electrons and
ions are out of temperature equilibrium. Since radiative shocks are the
main application for CRASH, we have
implemented a separate electron pressure equation with the electron thermal
heat conduction. For the electron heat conduction, we have added the option of
a flux limiter to limit the thermal flux with the free-streaming heat flux.

The multi-material radiation hydrodynamic equations are solved with an
operator split method that consists of three substeps: (1) solving the
hydrodynamic equations with standard finite volume shock-capturing schemes,
(2) the linear advection of the radiation in frequency-logarithm space, and
(3) the implicit solution
of the radiation, heat conduction, and energy exchanges. For the implicit
solver, standard Krylov solvers are used together with a Block Incomplete
Lower-Upper decomposition (BILU) preconditioner. This preconditioner scales
well up to 500 or 1000 processors. For future work, we may explore for the
implicit multigroup diffusion a multi-level preconditioner to better scale the
radiation solver beyond 1000 processors.

We have presented a suite of verification tests that benchmark the performance.
These tests verify the correctness and accuracy of the implementation for the
gray and multigroup radiation diffusion algorithm and the heat conduction in
1D, 2D, and 3D slab and 2D $rz$ geometry. To demonstrate the full capability
of the implementation, we have presented a 3D multi-material simulation of a
radiative shock wave propagating through an elliptic nozzle. This
configuration will be used in future validation studies.

Since this radiation transfer library is an extension of the BATS-R-US code,
the implementation is readily available for MHD simulations as well. This
allows for validation studies of the radiation MHD implementation using
laboratory-astrophysics experiments or for the simulations of astrophysical
plasmas.

\acknowledgments
This work was funded by the Predictive Sciences Academic Alliances Program
in DOE/NNSA-ASC via grant DEFC52-08NA28616 and by the University of Michigan.
The simulations were performed on the NASA Advanced Supercomputing system
Pleiades.

\appendix
\section{DISCRETIZATION OF THE DIFFUSION OPERATOR AT RESOLUTION CHANGES}
\label{sec:reschange}

\begin{figure}
\plotone{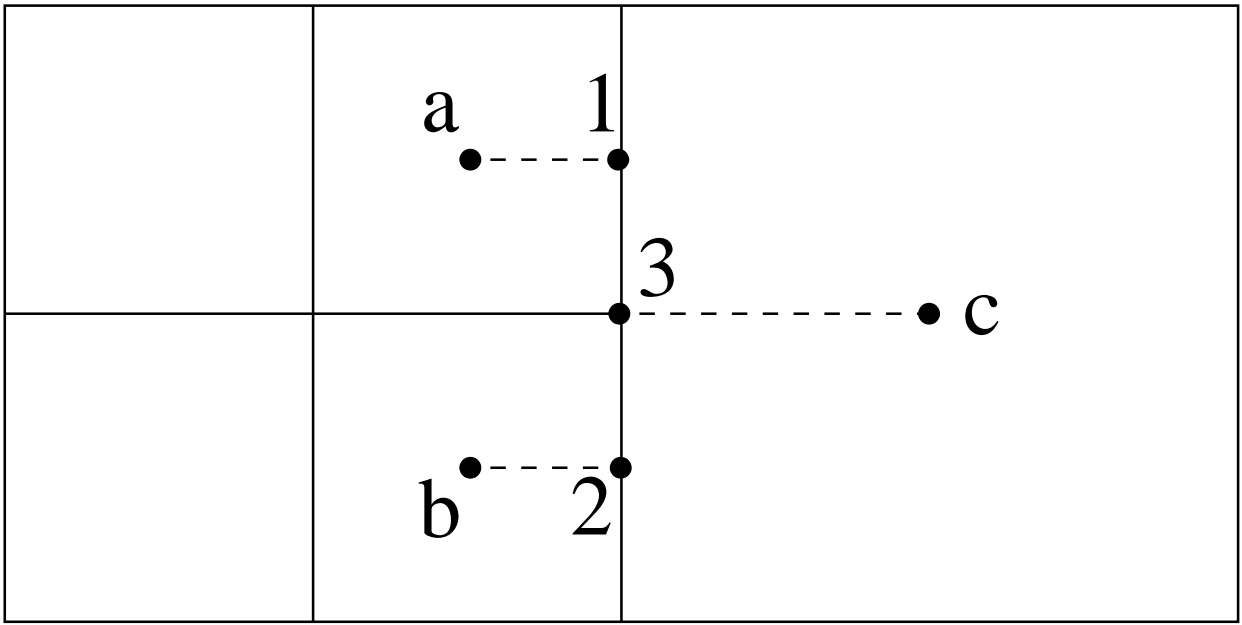}
\caption{Cell and face centers at the adaptive interface in 2D.}
\label{fig:grid}
\end{figure}

In Sections \ref{sec:unsplit} and \ref{sec:split}, the diffusion operator
is discretized on a uniform mesh with a standard
finite volume method in combination with a central difference approximation
for the gradient in the flux calculation as in equation
(\ref{eq:diffusionflux}). The diffusion coefficient that is needed at the
face is obtained by simple arithmetic averaging of the
left and right cell center diffusion coefficients. The generalization to
resolution changes as in Figure \ref{fig:grid} is less straightforward.
In the following, we will denote the fine cell centers by $a$ and $b$, the
coarse cell center by $c$. The flux densities at the resolution changes
in the direction orthogonal to the interface are denoted by $F_1$ and $F_2$
at the fine faces, and $F_3$ at the coarse face.

In \citet{edwards1996}, a strategy was developed to discretize the diffusion
operator on an adaptive mesh in the context of reservoir simulations. The
main ingredients of the method are (1) require the continuity of the flux
at the resolution change in the strong sense, i.e. $F_1=F_2=F_3$, and (2)
discretize the gradient in the diffusion flux by a one-sided difference.
An expression was found for the diffusion flux ${\bf F} = -D\nabla E$ in
which the diffusion coefficient is replaced by a weighted harmonic average of
the cell centered values $D_a$, $D_b$, $D_c$. In \citet{gittings2008},
it was argued that this discretization does not properly propagate the
self-similar Marshak waves of the radiation diffusion model, unless
the cell centered diffusion coefficients are calculated on a common face
temperature.

In the code discussed in this paper, we follow a different approach that
replaces the harmonic
average of the diffusion coefficient in \cite{edwards1996} by an arithmetic
average and obtain for the flux densities normal to the resolution change
interface
\begin{equation}
  F_1 = F_2 = F_3 = -\frac{2D}{3\Delta x}\left[ E_c - (E_a + E_b)/2 \right],
\end{equation}
where $\Delta x$ is the fine cell size and the diffusion coefficient at the
face is averaged as
\begin{equation}
  D_3 = \left[ D_c + (D_a+D_b) \right] /3.
\end{equation}
We demonstrated with verification tests including those discussed above that
this change produces properly
propagating radiative precursor and shock fronts. Generalizations to 1D and
3D are straightforward.

\section{RZ-GEOMETRY}\label{sec:rz}

Incorporating the $rz$-geometry in a finite volume formulation is as
follows: the radial cell face area and the cell volume must be made
proportional to the distance $r$ from the symmetry axis. In addition, the
$r$ component of the momentum equation ($\ref{eq:momentum}$) is modified as
\begin{equation}
  \frac{\partial\rho u_r}{\partial t} + \nabla\cdot\left[
    \rho\bu u_r + {\bf \hat r}(p+p_r)\right] = \frac{p+p_r}{r},
\end{equation}
where ${\bf \hat r}$ is the unit vector in the $r$ direction and 
$u_r=\bu\cdot{\bf \hat r}$. This correction reflects the fact that the pressure
term is a gradient, not a divergence.

\clearpage

\begin{deluxetable}{ r | c | l }
  \tablecaption{Quantities stored in the EOS tables as a function of
    $\log T_e \left[ \mbox{eV} \right]$ and
    $\log n_a \left[ \mbox{m}^{-3} \right]$.\label{table:eos}}
  \startdata
      {\bf quantity} & {\bf stored quantity} & {\bf units} \\
      \hline
      total pressure $p$ &$p/n_a$ & eV \\
      total internal energy density $E$ & $E/n_a$ & eV \\
      electron pressure $p_e$ & $p_e/n_a$ & eV \\
      electron internal energy density $E_e$ & $E_e/n_a$ & eV \\
      specific heat $C_{V}$ & $C_{V}/(n_ak_B)$ & \\
      electron specific heat $C_{Ve}$ & $C_{Ve}/(n_ak_B)$ & \\
      speed of sound gamma & $\gamma_{S}$ &  \\
      electron speed of sound gamma & $\gamma_{S_e}$ &  \\
      inverse of ion-electron interaction time & $1/\tau_{ie}$ & s$^{-1}$ \\
      electron conductivity & $C_e$ & J\,m$^{-1}$\,s$^{-1}$\,K$^{-1}$ \\
      mean ionization & $\overline{Z}$ & \\
      mean square ionization & $\overline{Z^2}$ &
      \enddata
\end{deluxetable}

\clearpage

\begin{deluxetable}{ r | c | l }
  \tablecaption{Quantities stored in the opacity tables as a function of
    $\log\rho \left[ \mbox{kg m}^{-3} \right]$ and
    $\log T_e \left[ \mbox{eV} \right]$.\label{table:opacity}}
  \startdata
      {\bf quantity} & {\bf symbol} & {\bf units} \\
      \hline
      specific group Rosseland mean opacities & $\kappa_{Rg}/\rho$ &
      kg$^{-1}$\,m$^2$ \\
      specific group Planck mean opacities & $\kappa_{Pg}/\rho$ &
      kg$^{-1}$\,m$^2$\\
      \enddata
\end{deluxetable}

\clearpage

\clearpage

\clearpage

\clearpage

\clearpage

\clearpage

\clearpage

\clearpage

\clearpage

\clearpage

\clearpage

\clearpage

\clearpage

\clearpage

\clearpage

\clearpage

\clearpage

\clearpage

\clearpage

\end{document}